\documentclass[11pt]{article}
\title{\bf Lagrangian perfect fluids and black hole mechanics}
\author{ Vivek Iyer\\Enrico Fermi Institute\\
University of Chicago\\5640 S. Ellis Ave.\\Chicago, IL 60637}
\date{\today} 

\usepackage{latexsym}
\begin{document}

\maketitle

\newcommand{\be}{\begin{equation}}
\newcommand{\ee}{\end{equation}}
\newcommand{\bea}{\begin{eqnarray}}
\newcommand{\eea}{\end{eqnarray} }
\newcommand{\beas}{\begin{eqnarray*}}
\newcommand{\eeas}{\end{eqnarray*} }
\newcommand{\bdm}{\begin{displaymath}}
\newcommand{\edm}{\end{displaymath} }

\newcommand{\del}{{\bf \nabla}}
\newcommand{\dee}{\partial}
\newcommand{\eps}{\mbox{{\boldmath $\epsilon$}}}
\newcommand{\Om}{\Omega}
\newcommand{\om}{\mbox{{\boldmath $\omega$}}}
\newcommand{\blam}{\mbox{{\boldmath $\lambda$}}}
\newcommand{\kt}{\tilde{\xi}}
\newcommand{\de}{\delta}
\newcommand{\kap}{\kappa}
\newcommand{\cd}{\cdot}
\newcommand{\Liek}{{\cal L}_\xi}
\newcommand{\Liekt}{{\cal L}_{\kt}}
\newcommand{\Liet}{{\cal L}_t}
\newcommand{\Liev}{{\cal L}_v}
\newcommand{\syp}{{\bf \Theta}}
\newcommand{\bB}{{\bf B}}
\newcommand{\bC}{{\bf C}}
\newcommand{\bS}{{\bf S}}
\newcommand{\bs}{{\bf s}}
\newcommand{\bm}{{\bf m}}
\newcommand{\bT}{{\bf T}}
\newcommand{\bt}{{\bf t}}
\newcommand{\bU}{{\bf U}}
\newcommand{\bL}{{\bf L}}
\newcommand{\bJ}{{\bf J}}
\newcommand{\bK}{{\bf K}}
\newcommand{\bQ}{{\bf Q}}
\newcommand{\bE}{{\bf E}}
\newcommand{\ble}{{\bf e}}
\newcommand{\bW}{{\bf W}}
\newcommand{\bX}{{\bf X}}
\newcommand{\bXt}{{\bf \tilde{X}}}
\newcommand{\bY}{{\bf Y}}
\newcommand{\bZ}{{\bf Z}}
\newcommand{\bff}{{\bf f}}
\newcommand{\bmu}{\mbox{\boldmath $\mu$}}
\newcommand{\bnu}{\mbox{\boldmath $\nu$}}
\newcommand{\scri}{{\cal I}}

\begin{abstract}
The first law of black hole mechanics (in the form derived by Wald), 
is expressed in terms of integrals over {\em surfaces}, at
the horizon and spatial infinity, of a stationary, axisymmetric black hole, 
in a diffeomorphism invariant Lagrangian theory of gravity.
The original statement of the first law given by Bardeen, Carter and
Hawking for an Einstein-perfect fluid system contained, in addition, 
{\em volume} integrals of the fluid fields, over a spacelike 
slice stretching between these two surfaces. 
One would expect that Wald's methods, applied to a 
Lagrangian Einstein-perfect fluid formulation, would convert 
these terms to surface integrals. However, because the fields 
appearing in the Lagrangian of a gravitating perfect fluid 
are typically nonstationary, (even in a stationary black hole-perfect 
fluid spacetime) a direct application of these  
methods generally yields restricted results. 
We therefore first approach the problem of incorporating 
general nonstationary matter 
fields into Wald's analysis, and derive a first law-like 
relation for an arbitrary Lagrangian metric 
theory of gravity coupled to {\em arbitrary} Lagrangian matter fields, 
requiring only that the {\em metric} field be stationary. This relation
includes a volume integral of
matter fields over a spacelike slice between the black hole horizon and 
spatial infinity, and reduces to the first law originally derived 
by Bardeen, Carter and Hawking when the theory is general relativity 
coupled to a perfect fluid. We then turn to 
consider a specific Lagrangian formulation for an isentropic
perfect fluid given by Carter, and directly apply Wald's analysis, assuming 
that both the metric and fluid fields are stationary and axisymmetric in the
black hole spacetime. The first law we derive contains only 
surface integrals at the black hole horizon and spatial infinity, 
but the assumptions of 
stationarity and axisymmetry of the fluid fields make this relation much more 
restrictive in its allowed fluid configurations and perturbations than 
that given by Bardeen, Carter and Hawking. 
In the Appendix, we use the symplectic structure of the
Einstein-perfect fluid system to derive a conserved current for perturbations
of this system: this current reduces to  
one derived {\em ab initio} for this system by Chandrasekhar and Ferrari.
\end{abstract}

\section{Introduction}
The first law of black hole mechanics as stated by Bardeen, Carter and 
Hawking \cite{bch} relates small changes in the mass of a stationary,
axisymmetric black hole to small changes in its horizon surface area, angular 
momentum and the properties of a stationary perfect fluid that might 
surround it: one first fixes a stationary axisymmetric
Einstein-perfect fluid black hole solution 
with stationary killing field $\xi^a$ (with asymptotically unit norm)
and axial killing field $\varphi^a$ (with closed orbits). One then 
defines $\de$ to be an infinitessimal
perturbation to a nearby stationary axisymmetric solution;
then the first law in \cite{bch} is 
\be
\de M = \frac{\kappa}{8\pi}\de A + \Om_H \de J_H - \int_{\Sigma}\mu'|v|
\de N_{abc} + \int_{\Sigma}\Om\de J_{abc}+\int_{\Sigma}T|v|\de S_{abc} ,
\label{bchfl}
\ee
where the spacetime is characterised by an ADM mass, $M$, and the black
hole by its horizon surface area, $A$, surface gravity, $\kappa$,
angular velocity, $\Omega_H$, and angular momentum $J_H$ (measured
at the horizon).
The fields associated to the perfect fluid are its four velocity, $U^a$
(which here is taken to be of the form
$U^a = v^a/|v|$, where $v^a = \xi^a +\Omega\varphi^a$, for some 
(generally non-constant) $\Omega$),
the chemical potential $\mu'$, the temperature $T$, stress-energy $T^{ab}$,
and number and entropy densities $n$ and $S$.
The three-forms $N_{abc} = nU^d\eps_{abcd}$, $J_{abc} =
{T^d}_e\varphi^e\eps_{dabc}$, and $S_{abc} = S U^d\eps_{dabc}$ represent
the fluid number density, angular momentum density and entropy density
on a spacelike 3-surface, $\Sigma$, that has boundaries at the black hole
horizon and  the two-sphere at spatial infinity. 
We have also set $\eps_{abcd}$ to be the canonical volume element 
on spacetime.

Considerable effort has been spent on weakening the assumptions made
in (\ref{bchfl}) on the background fields and their perturbations.
For instance, consider an arbitrary diffeomorphism invariant Lagrangian
theory with both metric and matter fields, and let the theory
possess stationary, axisymmetric black hole solutions, which are 
asymptotically flat, and have a bifurcate killing horizon
(for an explanation of these terms see \cite{Wald,WaldGR}). 
Then it was shown \cite{Wald,IW1}, providing the metric 
and matter fields appearing in the Lagrangian were stationary and 
axisymmetric in the black hole background, that there existed a first 
law of black hole mechanics in a form only 
involving {\em surface integrals} on the sphere at spatial infinity and the
bifurcation sphere of the black hole horizon. 
Namely, given the Lagrangian for the theory, 
one could algorithmically 
define integrals $\cal E$ and $\cal J$
over the sphere at spatial infinity, and $\cal S$ over the bifurcation
sphere, satisfying the following identity:
\be
\de {\cal E} = \frac{\kappa}{2\pi}\de {\cal S} +\Omega_H\de{\cal J}.
\label{genfl}
\ee
(Here $\de$ denotes a  perturbation from the background black hole solution to 
{\em any} nearby solution.)
The quantity $\cal E$ was interpreted as the canonical energy of the black hole
system, $\cal J$ as the canonical angular momentum and $\cal S$ as the black 
hole entropy.

We might therefore expect that the volume integrals
in (\ref{bchfl}) involving the fluid can be converted to 
surface integrals in the form (\ref{genfl}),  
by choosing a suitable variational form for the Einstein-perfect fluid system
and using the methods of \cite{IW1}. In fact, we are unable to  
reproduce the first law (\ref{bchfl}) in a form only
containing surface integrals, using these methods;
the difficulty is that at least one of the fields appearing in each of the  
Lagrangian formulations for a perfect fluid (that we are aware of)
is generally non-stationary, even when
the fluid four velocity, number density, entropy, and functions of these
fields (which we refer to collectively 
as the {\em physical} fields), 
are stationary. Since the methods of \cite{IW1} require
that all fields appearing in the Lagrangian (which we refer to henceforth 
as the {\em dynamical fields}) are stationary and 
axisymmetric in the black hole background, the allowed background
solutions for the perfect fluid in the resulting first law are restricted. 

This paper gives two results in response to this problem: 
we first relax all explicit symmetry assumptions on matter fields 
appearing in the Lagrangian, and 
find the consequence for the first law given in \cite{IW1}.
We also attempt to generate a first law of the form
(\ref{genfl}) by a careful choice of an existing Lagrangian formulation for 
gravity coupled to a perfect fluid, directly using the methods of \cite{IW1}.

In section (2) we consider an arbitrary Lagrangian theory of 
gravity coupled to {\em arbitrary} matter fields, assuming only that the
metric is stationary and axisymmetric in the black hole background,
but making no such assumptions about the matter dynamical fields.
We then modify the methods 
of \cite{IW1} to generate a perturbative relation, 
but instead of attempting to express the matter contribution 
to the first law (\ref{genfl}) via surface integrals, we 
leave it instead as a volume integral over a  
hypersurface, $\Sigma$, joining the bifurcation sphere to the sphere at
spatial infinity. In restricted cases (which
we explain later) we can motivate an independent measurement 
of the ``vacuum" black hole mass, $M_g$. In these cases we can also 
define quantities which resemble the ``vacuum" black hole entropy,
$S_g$, and angular momentum, ${J_g}_H$, and having done so 
our perturbative relation takes the form
\be
\de M_g = \frac{\kappa}{2\pi}\de S_g + \Omega_H \de {J_g}_H,
+\int_{\Sigma}
\frac{1}{2}\xi\cdot\eps \; T^{ab} \; \de g_{ab}- \de(\eps\cdot\; T\cdot \xi),
\label{volfl}
\ee
where $T^{ab}$ is the stress-energy of the matter fields. We will see
that this relation defines a black hole entropy, $S_g$, 
which is in general 
not the black hole entropy defined in \cite{IW1}: however, in special 
cases the interpretation of $S_g$ as black hole entropy can be appropriate 
(for instance, as we show in section (4), this relation reduces to 
(\ref{bchfl}) when the gravitational theory is chosen to be general 
relativity, and the matter source is chosen to be a perfect fluid).
Our result differs from a similar relation presented by Schutz and
Sorkin \cite{ss}, in that they conjectured, but did not explicitly 
include the black hole entropy and angular momentum
boundary terms, and so did not explicitly generalise the 
full form of (\ref{bchfl}). In addition, as we shall explain, 
the definition of our ``Noether current" (involved in the intermediate
calculations) is both less ambiguous 
than that presented by Schutz and Sorkin \cite{ss} 
and more general than the definition given by 
Sorkin \cite{Sorkin92}.  The range of theories in 
which our methods are well defined is therefore larger than those addressed
by their methods.

In section (3) we define a gravitating perfect fluid
and review some variational principles for it:
Schutz's ``velocity-potential" formulation \cite{SchF}, which uses
the dynamical fields $(\phi,\alpha,\beta,\theta,\sigma)$ to define the 
product of the (physical) specific inertial mass and four velocity - 
$\mu U_a \equiv \del_a\Phi+\alpha\del_a\beta+\theta\del_a \sigma,$
and Carter's more recent ``axionic vorticity" formulation \cite{carter}
for an isentropic perfect fluid, 
which uses a dynamical field $b_{ab}$ to define the 
number current $N_{abc}$ (given in (\ref{bchfl})) via 
$N_{abc}\equiv 3\del_{[a}b_{bc]}$, and the dynamical fields 
$\chi^{\pm}$ to define the fluid vorticity via $2\del_{[a}\mu U_{b]}\equiv
2\del_{[a} \chi^+\del_{b]}\chi^-$.

In section (4) we present two forms of the first law for the 
Einstein-perfect fluid system. The first form is derived from the 
relation (\ref{volfl}) and is the same as (\ref{bchfl}),
with the exception that $\de$ is now allowed to be a perturbation from 
the (stationary axisymmetric) background to an arbitrary nearby solution. 
(Note that this form of the first law contains volume integrals.)
It is of also interest to know if we can construct {\em any} form of the first
law with perfect fluids only involving surface integrals; in fact,
by directly applying the methods of \cite{IW1} for a metric 
theory of gravity coupled to a perfect fluid described using Carter's 
variational principle, (with the potential $b_{ab}$ for $N_{abc}$, and
$\chi^{\pm}$ for $\omega_{ab}$) we can derive a first law of the form  
\be
\de M +\mu_{\infty}\de\int_{S_{\infty}} b_{qr}
-\mu_{\infty}\de\int_{\cal H}b_{qr}=\frac{\kappa}{8\pi}\de A 
-\Omega_H\de J_H + \int_{\cal H} X_{qr} - \int_{S^{\infty}} X_{qr}, 
\label{intfl}
\ee
where $M$ is the ADM mass, $A$ is the black hole surface area, $J_H$
is the black hole angular momentum appearing in (\ref{bchfl}), 
$X_{qr}$ is the two-form 
$2\xi^pb_{p[q}[\de (\mu U_{r]}) - \del_{r]}\chi^-\de\chi^++\del_{r]}\chi^+
\de \chi^-]$, and we
have written $S^{\infty}$ and $\cal H$ for the sphere at spatial infinity
and the bifurcation sphere, respectively. 
We will see that this first law 
is more restrictive than (\ref{bchfl}), but 
it is the only non-trivial rule of the type (\ref{genfl}) 
involving a perfect fluid that we can currently construct. 

In the Appendix we evaluate the symplectic form of
the Einstein-perfect fluid system, using the variational formulation
given by Schutz \cite{SchF} for the perfect fluid. The symplectic
form is dual to a generally conserved current, 
quadratic in the field perturbations \cite{LW}. 
We find (in parallel with Burnett and Wald's calculation for the 
Einstein-Maxwell system \cite{BW}) that this conserved current reduces to a 
current previously derived {\em ab initio} by Chandrasekhar and Ferrari 
\cite{CF} for the polar perturbations of a static axisymmetric black hole.

\section{A perturbative relation for black hole mechanics with
non-stationary matter fields}
In this section we give a perturbative relation that resembles the first 
law of black hole mechanics, for an arbitrary theory of 
gravity with a diffeomorphism invariant Lagrangian. 
We assume the theory possesses black hole solutions
in which the metric is stationary and axisymmetric, but place no 
restrictions on 
the other fields appearing in the Lagrangian (we refer to these fields
collectively as the {\em dynamical} fields). The motivation for this is, 
as we have indicated, that variational formulations for gravitating
Einstein-perfect fluid systems have fluid dynamical fields 
which are nonstationary even when the fluid's {\em physical} fields
(the four-velocity, number density and  entropy) are stationary 
and axisymmetric. We first make some necessary definitions related to the
the symplectic structure of a diffeomorphism invariant Lagrangian
theory. These are explained in detail in \cite{IW1}; here we 
merely state (and, in one case, refine) the relevant 
definitions and results. In the following 
we often use bold face type to denote differential forms on spacetime, 
suppressing their indices when convenient.
 
\subsection{Some Preliminaries}
All theories we consider arise from a Lagrangian, which is taken to be a  
diffeomorphism invariant four-form on spacetime, 
dependent on the metric, $g_{ab}$, and some arbitrary set of matter 
fields, $\psi$. (We collectively refer to all the dynamical fields by 
$\phi$.) By this we mean that the Lagrangian has the functional dependence
\be
\bL = \bL(g_{ab}, R_{abcd}, \del R_{abcd},\ldots,(\del)^p R_{abcd},
\psi,\del \psi,\ldots,(\del)^q \psi),
\label{Lfirst}
\ee
(here multiple derivatives appearing in the above expression are
assumed to be symmetrised - see \cite{IW1} for further discussion about 
this dependence). In particular
we require that every field appearing in the Lagrangian give rise to an
equation of motion (there are no ``background" fields).
The variation of the Lagrangian defines these equations, $\bE =0$, along with
the symplectic potential $\syp$, by
\be
\de \bL = \bE \de \phi +  d\syp(\phi,\de\phi).
\label{delfirst}
\ee 
(Here $\syp(\phi,\de\phi)$ is a linear differential operator
in the field variations $\de\phi$. Because the Lagrangian is only defined
up to the addition of an exact form, $\bL\rightarrow\bL+d\bmu$, 
the symplectic potential is only defined up to the following
terms:$\syp(\phi,\de\phi)\rightarrow\syp(\phi,\de\phi)+d\bY(\phi,\de\phi)+
\de\bmu(\phi)$, where $\bY$ and $\bmu$ are covariant forms with the same type
of functional dependence as $\syp$ and $\bL$, respectively. These ambiguities
were discussed in \cite{IW1}.)

Now fix a smooth vector field, $\xi^a$, on spacetime. Then the Noether current
$\bJ[\xi]$ associated to $\xi^a$ is a three-form defined by 
\be
\bJ[\xi]\equiv \syp(\phi,\Liek\phi)-\xi\cdot\bL 
\label{J}
\ee
where the centred dot denotes contraction of the vector into the first
index of the form. This Noether current can be seen \cite{IW1} 
to obey the following identity:
\be
d\bJ[\xi] = -\bE\Liek \phi,
\label{Jclose}
\ee
which we now use to further elucidate its structure.
(Although they appear in a different context, the calculations below 
have the same flavour as those in the Appendix of \cite{IW1}.)

{\bf Lemma 1:} Fix $\bL$ to be the Lagrangian of a 
diffeomorphism invariant theory of gravity and matter fields, 
with equation of motion $\bE=0$ as given in (\ref{delfirst}).  
Without loss of generality, label each dynamical 
field $\phi$ by $i$, and give each field $u_i$ upper, and $d_i$ lower indices:
also label the equations of motion for each field similarly, so that the 
equation of motion term in (\ref{delfirst}) becomes
\be
\bE\de\phi= \eps
{{E_{\phi_i}}_{b_1 \cdots {b_u}_i}}^{a_1 \cdots {a_d}_i}
\de{{\phi_i}^{b_1\cdots b_{u_i}}}_{a_1\cdots a_{d_i}}.
\label{philabel}
\ee
Then for any smooth field $\xi^a$ there exists a two-form,
$\bQ[\xi]$, called the Noether charge associated to $\xi^a$
(which is local in the dynamical fields and $\xi^a$), such that 
the Noether current $\bJ[\xi]$, defined in (\ref{J}), can be written
\be
\bJ[\xi]=-(\eps\cdot E\cdot\phi\cdot\xi) +d\bQ[\xi],
\label{eEx}
\ee
where we define the three-form
\bea
(\eps\cdot E\cdot\phi\cdot\xi)_{abc}&\equiv&
\eps_{eabc}\sum_i
{{E_{\phi_i}}_{b_1 \cdots {b_u}_i}}^{a_1 \cdots {a_d}_i}(
-{{\phi_i}^{e \cdots {b_u}_i}}_{a_1 \cdots {a_d}_i}\de^{b_1}_p
\ldots -{{\phi_i}^{b_1 \cdots e}}_{a_1 \cdots {a_d}_i}\de^{b_{u_i}}_p
\nonumber\\
&&+{{\phi_i}^{b_1\cdots {b_u}_i}}_{p \cdots {a_d}_i}\de^{e}_{a_1}
\ldots +{{\phi_i}^{b_1 \cdots b_{u_i}}}_{a_1 \cdots p}\de^e_{a_{d_i}})\xi^p
\label{eExdef}
\eea

{\bf Proof:}\\
For clarity, we first consider the case where the 
metric is the only dynamical field: $\phi\rightarrow g_{ab}$. Then 
setting the metric field equations $\bE^{ab}=\eps E^{ab}$, Eq.(\ref{Jclose})
reads
\bea
d\bJ[\xi]&=&-2\eps E^{ab}_g\del_a\xi_b\nonumber\\
&=&-2\eps\del_a(E^{ab}_g\xi_b)+2\eps\del_a(E^{ab}_g)\xi_b.
\label{Jmanip}
\eea
Therefore setting $(\eps\cdot E_g\cdot\xi)_{abc}\equiv\eps_{dabc}E^{de}_g\xi_e$,
we have 
\be
d(\bJ[\xi]+2\eps\cdot E_g\cdot\xi)=2\eps\del_a(E_g^{ab})\xi_b,
\label{dJE}
\ee
which shows that the right side of (\ref{dJE}) is both linear in $\xi^a$, and
exact for all $\xi^a$. The results of \cite{onclosed} 
now imply that the right side must vanish identically - and so 
$\del_a E^{ab}=0$. This in turn implies that the left side of (\ref{dJE})
must be an identically closed three-form, which (using the results of 
\cite{onclosed} again) implies the existence of a two-form, $\bQ[\xi]$,
local in the dynamical fields and $\xi^a$, such that
\be
\bJ[\xi]+2\eps\cdot E_g\cdot\xi =d\bQ[\xi].
\ee
We define $\bQ[\xi]$, the Noether charge associated to $\xi^a$,
as any two form which is local in the dynamical fields and $\xi^a$, 
and satisfies this relation.

We can also perform this analysis for $\bL$ with
the general dependence (\ref{Lfirst}). With the labels for each field and
its equation of motion given in (\ref{philabel}),
the first equation in (\ref{Jmanip}) becomes
\be
d\bJ[\xi]= -\eps\sum_i
{{E_{\phi_i}}_{b_1 \cdots {b_u}_i}}^{a_1 \cdots {a_d}_i}\Liek
{{\phi_i}^{b_1\cdots b_{u_i}}}_{a_1 \cdots {a_d}_i},
\label{dJgen}
\ee
which, through a similar manipulation to (\ref{Jmanip}) leads to the structure
for $\bJ[\xi]$ and the definition of the Noether charge,
$\bQ[\xi]$, in (\ref{eEx}). $\Box$

The Noether charge was defined in \cite{IW1} only when $\bE=0$, via
$\bJ[\xi]=d\bQ[\xi]$. This left open the definition 
of $\bQ[\xi]$ when $\bE\ne 0$. 
In the Appendix of \cite{IW2}, however, it was shown that $\bQ[\xi]$ could 
be defined when $\bE\ne 0$, such that there existed forms ${\bf C}_a$
with $\bJ[\xi]-d\bQ[\xi]={\bf C}_a\xi^a$, 
and where the ${\bf C}_a$ vanished when $\bE=0$. 
At that time it was not known whether
$\bQ[\xi]$ was uniquely defined this way, nor was the explicit form  
of ${\bf C}_a$ specified. We have given this explicit form 
in (\ref{eExdef}). Moreover, the above analysis uniquely 
defines the Noether charge via (\ref{eEx}), without imposing
the field equations, up to the following ambiguities (which were discussed
in detail in \cite{IW1}):
The ambiguity in $\syp$ described after Eq. (\ref{delfirst})
means that $\bJ[\xi]$ is only defined up to the following terms:
$\bJ[\xi]\rightarrow\bJ[\xi]+d(\bY(\phi,\Liek\phi)-\xi\cdot\bmu)$, 
and so the ambiguity in $\bQ[\xi]$ is $\bQ[\xi]\rightarrow\bQ[\xi]
+\bY(\phi,\Liek\phi)-\xi\cdot\bmu$. These ambiguities will not affect
the results stated in the following sections.

We now define the symplectic current, $\om(\phi,\de_1\phi,\de_2\phi)$, (a 
three-form on spacetime) by:
\be
\om(\phi,\de_1\phi,\de_2\phi)\equiv\de_2\syp(\phi,\de_1 \phi) -\de_1\syp(\phi,
\de_2 \phi).
\label{omega}
\ee
Note that $\om$ is a function of an unperturbed set of 
fields, $\phi$, and is bilinear and skew in pairs of variations 
$(\de_1\phi,\de_2\phi)$. 
It can be shown (see \cite{LW}) that this three-form is closed when
$\phi$ is a solution of the field equations and 
$\de_1\phi$ and $\de_2\phi$ are solutions of the linearised 
equations of motion (In the Appendix we examine this closed form - it 
is dual to a conserved vector field, which we evaluate for 
perturbations of an Einstein-perfect fluid system). 
Moreover, if we let $\xi^a$ be a smooth vector field, 
set $\de_1\phi = \Liek \phi$ and let $\de_2\phi=\de\phi$ be a 
variation to a nearby solution (with $\de\xi^a=0$), then
$\om(\phi,\Liek\phi,\de\phi)$ can be shown \cite{IW1} to be exact: 
\be
\om(\phi,\Liek\phi,\de\phi)=
d[\de\bQ[\xi]-\xi\cdot\syp(\phi,\de\phi)].
\label{exom}
\ee

Now fix a black hole spacetime with a stationary and axisymmetric metric,
for the theory given by the Lagrangian in (\ref{Lfirst}); let
the stationary killing field with unit norm at spatial infinity be $\xi^a$
and the axial killing field (with closed orbits) be $\varphi^a$. 
Let the black hole have a bifurcate killing horizon, with bifurcation sphere
$\cal H$, and let it be asymptotically flat, with the two-sphere at spatial 
infinity $S^{\infty}$. Let $\Sigma$ be a three-surface with these
two boundaries, and set $\de\phi$ to be an arbitrary perturbation of the 
background which satisfies the linearised equations.
Then the first law of black hole mechanics as 
stated in \cite{IW1} is an interpretation of the identity  
\be
\int_{\Sigma}\om(\phi,\Liek\phi,\de\phi)=
\int_{S_{\infty}}\de\bQ[\xi]-\xi\cdot\syp(\phi,\de\phi)
-\int_{\cal H}\de\bQ[\xi]-\xi\cdot\syp(\phi,\de\phi)
\label{wfl}
\ee
(which arises from integrating (\ref{exom}) over $\Sigma$). When  
$\xi^a$ Lie derives {\em all} the dynamical fields in the background,
the left side of (\ref{wfl}) vanishes, and one is left with a 
relation between surface integrals on the boundaries of 
$\Sigma$, which can be shown to be of the form (\ref{genfl}). 
In section (4) we present an explicit 
Lagrangian for the Einstein-perfect fluid system, 
and, assuming that all dynamical
fields are stationary and axisymmetric, compute the surface terms 
arising from this Lagrangian.

\subsection{The perturbative identity}
Having stated these necessary definitions we turn to construct our perturbative
identity. We start by decomposing the Lagrangian 
$\bL$ into a part $\bL_g$, depending on the metric, $g_{ab}$,
(which is assumed to be 
stationary and axisymmetric in the black hole background),
and a part $\bL_m$, dependent on both the metric 
and a set of matter fields, $\psi$, (on which we place no restrictions): 
\bea
\bL&=&\bL_g(g_{ab}, R_{abcd},\del R_{abcd},\ldots,(\del)^p R_{abcd})\nonumber\\
&&+\bL_m(\psi,\del \psi,\ldots,(\del)^q \psi, g_{ab}, R_{abcd}, 
\del R_{abcd},\ldots,(\del)^r R_{abcd}).
\label{Lgen}
\eea
Since this breakup only requires that $\bL_g$ be independent of any
matter fields, it is very non-unique, and in general  we have no method of 
controlling the ambiguity
\bea
\bL_g&\rightarrow&\bL_g + \blam,\nonumber\\
\bL_m&\rightarrow&\bL_m - \blam.
\label{Lambig}
\eea
where $\blam=\blam(g_{ab}, R_{abcd}, \del R_{abcd},\ldots,(\del)^s R_{abcd})$.

The variation of the Lagrangian yields equations of motion for the metric, 
$\bE^{ab}_g=0$, and matter fields, $\bE_m=0$, via
\be
\de \bL = \bE^{ab}_g \de g_{ab} + \bE_m \de \psi + d\syp(\phi,\de\phi).
\label{deL}
\ee
For convenience we set $\bE^{ab}_g=\eps E^{ab}_g$ and $\bE_m=\eps E_m$.
As discussed above we can compute $\bJ[\xi]$ (defined by (\ref{J})),
and define $\bQ[\xi]$, for the theory described by (\ref{Lgen}): 
it must have the form given in (\ref{eEx}):
\be
\bJ[\xi]=-2\eps\cdot E_g\cdot\xi - \eps\cdot E_m\cdot\psi\cdot\xi + d\bQ[\xi]
\label{Js}
\ee
(the factor of two between the terms with equations of motion here is
purely a matter of convention).
 
We can also use the individual Lagrangians
$\bL_g$ and $\bL_m$ to define the stress-energy tensor $T^{ab}$,
and symplectic potentials ${\syp_g}(g,\de g)$ and ${\syp_m}(\phi,\de\phi)$:
\bea
\de\bL_g&=&\bE^{'ab} \de g_{ab} + d{\syp_g}(g,\de g) \nonumber\\
\de\bL_m&=&\bE_m\de \psi+\eps\frac{1}{2}T^{ab}\de g_{ab}
+d{\syp_m}(\phi,\de\phi).
\label{deLm}
\eea
Clearly $\bE_g^{ab} = \bE_g^{'ab} + \eps\frac{1}{2}T^{ab}$, and 
up to the ambiguities present in the symplectic potentials, we also have 
$\syp = {\syp_g} + {\syp_m}$.  
Similarly, if we define the Noether currents for the individual Lagrangians by
\bea
{\bJ_g}[\xi] &\equiv&{\syp_g}(g,\Liek g)-\xi\cdot\bL_g,\nonumber\\
{\bJ_m}[\xi]&\equiv&{\syp_m}(\phi,\Liek\phi)-\xi\cdot\bL_m, 
\label{Jmat}
\eea
then it follows that  
\be
\bJ[\xi]={\bJ_g}[\xi]+{\bJ_m}[\xi].
\label{Jsum}
\ee
Now we impose the structure (\ref{eEx}) on each of ${\bJ_g}$ and ${\bJ_m}$,
in the process defining ${\bQ_g}$ and ${\bQ_m}$, 
which are the Noether charges in the theories arising from 
these Lagrangians: 
\bea
{\bJ_g}[\xi]&=&-2\eps\cdot E'_g\cdot\xi + d{\bQ_g}[\xi]\nonumber\\
{\bJ_m}[\xi]&=&-\eps\cdot E_m\cdot\psi\cdot\xi 
-\eps\cdot T \cdot\xi + d{\bQ_m}[\xi].
\label{Jstruct}
\eea

Finally we substitute (\ref{Jstruct}) into the right side of 
(\ref{Jsum}) and (\ref{Js}) into the left side, obtaining
\be
-2\eps\cdot E_g\cdot\xi -\eps\cdot E_m\cdot\psi\cdot\xi + d\bQ[\xi]=
-2\eps\cdot E'_g\cdot\xi 
-\eps\cdot T \cdot\xi 
-\eps\cdot E_m\cdot\psi\cdot\xi 
+ d{\bQ_g}[\xi]+ d{\bQ_m}[\xi].
\ee
All the terms involving equations of motion and stress-energy tensors
can be seen to cancel, and the resulting identity implies
\be
\bQ[\xi] = {\bQ_g}[\xi] + {\bQ_m}[\xi] +d\bZ,
\label{Qsum}
\ee
(where $\bZ$ is some arbitrary covariant one-form). We therefore have
a relation (independent of any field equations)
between the Noether charge, $\bQ$, of the full theory given
by $\bL$, and that of the ``pure gravity" theory ${\bQ_g}$, 
arising from $\bL_g$.
We are now ready to state the identity:

{\bf Lemma 2: }
Fix $\bL$, $\bL_g$ (the ``vacuum" Lagrangian) 
and $\bL_m$ (the ``matter" Lagrangian)
to be diffeomorphism invariant Lagrangians
related as given in (\ref{Lgen}) with the functional dependence shown there. 
Fix a smooth vector field $\xi^a$, let $\syp_g$ be defined by (\ref{deLm}),
let $\bQ_g[\xi]$ be the Noether charge defined by (\ref{Jstruct})
for the theory described by $\bL_g$, 
and let $T^{ab}$ be the stress-energy tensor of the matter fields
defined by (\ref{deLm}).
Now consider an asymptotically flat, 
stationary, axisymmetric black hole solution 
with bifurcate killing horizon, in the theory described by $\bL$, 
with stationary killing field $\xi^a$ (with unit norm at the sphere
$S^{\infty}$, at spatial infinity),
and axial killing field $\varphi^a$ (with closed orbits), so that $\xi^a$
and $\varphi^a$ Lie derive the metric but not necessarily the matter fields.
Let the horizon killing field (which vanishes on the bifurcation 
sphere $\cal H$) be given by $\chi^a = \xi^a+\Omega_H\varphi^a$, 
where $\Omega_H$ is a constant. Then for $\de$ a perturbation to an
arbitrary nearby solution, such that $\de\xi^a=0$, 
\be
\int_{S_{\infty}}\de{\bQ_g}[\xi]-\xi\cdot\syp_g= 
\int_{\cal H}\de{\bQ_g}[\chi]-\Omega_H\int_{\cal H}\de{\bQ_g}[\varphi]
+\int_{\Sigma}\frac{1}{2}\xi\cdot\eps\; T^{ab}\de g_{ab}
-\de(\eps\cdot\; T\cdot \xi). 
\label{prefl}
\ee

{\bf Proof:}\\
We evaluate the expression (\ref{wfl})
for the theory (\ref{Lgen}), where the background solution is 
a black hole with the symmetry and structure described above (\ref{prefl}), 
demanding that the metric be stationary and axisymmetric
in the background spacetime, but placing no restrictions on the matter fields.
In this case 
the integrand on the left side of (\ref{wfl}) is generally nonvanishing.
Assuming that the field equations hold in background for the 
matter fields, $E_m=0$, and that $\de\psi$
is a solution to the linearised matter equations of motion 
off this background ($\de E_m=0$), we find the left side of (\ref{wfl}) is
\bea
\om(\phi,\Liek\phi,\de\phi)&=&\de{\syp_g}(g,\Liek g)-\Liek\syp_g(g,\de g)
+\de{\syp_m}(\phi,\Liek \phi) -\Liek\syp_m(\phi,\de \phi)\nonumber\\
&=& \de{\syp_m}(\phi,\Liek\phi) -\Liek\syp_m(\phi,\de\phi)\nonumber\\
&=& \de(d{\bQ_m}[\xi] - \eps\cdot T\cdot \xi +\xi\cdot\bL_m) 
- \Liek{\syp_m}(\phi,\de\phi)\nonumber\\
&=& \de(d{\bQ_m}[\xi] - \eps\cdot T\cdot \xi +\xi\cdot\bL_m) 
- \xi\cdot d{\syp_m}(\phi,\de\phi) - d(\xi\cdot\syp_m(\phi,\de\phi))\nonumber\\
&=& d(\de{\bQ_m}[\xi] - \xi\cdot{\syp_m}(\phi,\de\phi))
- \de(\eps\cdot T\cdot\xi)+\frac{1}{2}\xi\cdot\eps\; T^{ab}\de g_{ab},
\label{omvol}
\eea
where we used the stationarity of $g_{ab}$ in the second line,
the expression (\ref{Jstruct}) for ${\bJ_m}$ in the third,  
the Lie derivative identity $\Liek\blam=\xi\cdot d\blam+d(\xi\cdot\blam)$
(which holds for an arbitrary form $\blam$) in the fourth line,
and the definition (\ref{deLm}) of ${\syp_m}$ and the stress-energy $T^{ab}$ 
in the fifth line. Now also assuming $E_g=\de E_g=0$, and  
substituting (\ref{omvol}) into the left side of (\ref{wfl}) yields
\be
\int_{\Sigma}d(\de{\bQ_m}[\xi]-\xi\cdot{\syp_m})+
\frac{1}{2}\xi\cdot\eps\; T^{ab}\de g_{ab}-\de(\eps\cdot\; T\cdot \xi) =
\int_{S_{\infty}}\de\bQ[\xi]-\xi\cdot\syp(\phi,\de\phi) -
\int_{\cal H}\de\bQ[\xi]-\xi\cdot\syp(\phi,\de\phi),
\ee
and so, cancelling the boundary terms $\de{\bQ_m}[\xi]-\xi\cdot{\syp_m}$ from 
both sides (and using (\ref{Qsum})) we get
\be
\int_{\Sigma}\frac{1}{2}\xi\cdot\eps\; T^{ab}\de g_{ab}
-\de(\eps\cdot\; T\cdot \xi) =
\int_{S_{\infty}}\de{\bQ_g}[\xi]-\xi\cdot{\syp_g}(g,\de g)-
\int_{\cal H}\de{\bQ_g}[\xi]-\xi\cdot{\syp_g}(g,\de g).
\label{interfl}
\ee

Now writing $\xi^a$ in terms of $\chi^a$ and $\varphi^a$ 
at the boundary $\cal H$ 
(and discarding terms which vanish as a result of 
the vanishing of $\chi^a$ at $\cal H$, or which vanish because 
$\varphi^a$ is tangent to $\cal H$) we get  
\be
\int_{\Sigma}\frac{1}{2}\xi\cdot\eps\; T^{ab}\de g_{ab}
-\de(\eps\cdot\; T\cdot \xi) =
\int_{S_{\infty}}\de{\bQ_g}[\xi]-\xi\cdot{\syp_g}(g,\de g)-
\int_{\cal H}\de{\bQ_g}[\chi]+\Omega_H\int_{\cal H}\de{\bQ_g}[\varphi],
\label{pprefl}
\ee
which is what we wished to show. $\Box$
 
The identity (\ref{prefl}) has physical significance when we
can interpret the surface integrals appearing there as (variations
of) the energy, entropy and angular momentum of the black hole. When
is this possible? If the theory had no matter fields 
then we could choose $\bL_m$ to vanish, and the terms involving 
$T^{ab}$ in (\ref{prefl}) would vanish (we could also choose other 
breakups of $\bL$, and we'll return to this shortly). 
In this case we'd have $\bL_g=\bL$, ${\syp_g}=\syp$, and ${\bQ_g} = \bQ$.
If in addition there existed a three-form $\bB$, (local in 
the dynamical fields, the flat metric $\eta_{ab}$, and its
associated derivative, $\dee$, at spatial infinity)
such that at spatial infinity, $\xi\cdot\syp(\phi,\de\phi)=\xi\cdot\de\bB$,
then (\ref{prefl}) can be written  
\be
\de {\cal E} = \frac{\kappa}{2\pi}\de {\cal S} +\Omega_H\de {\cal J}_H,
\label{genfl2}
\ee
where the varied quantities in (\ref{genfl2}) 
are defined below, and have well-known physical
interpretations \cite{IW1}. These are (i) The canonical energy of the
system, which we define as 
\be
{\cal E}\equiv\int_{S_{\infty}}\bQ[\xi]-\xi\cdot\bB,
\label{EH}
\ee
(ii) The entropy $\cal S$ of the black hole; 
by taking the functional derivative of the Lagrangian with respect to
the Riemann tensor (treated as an independent field) we know 
(setting $\eps_{ab}$ to be the binormal to the bifurcation sphere) that 
\be
\de \int_{\cal H}\bQ[\chi] =\frac{\kappa}{2\pi}\de {\cal S},  
\ee
where 
\be
{\cal S}\equiv-2\pi\int_{\cal H} 
\frac{\de \bL}{\de R_{abcd}}\eps_{ab}\eps_{cd},
\label{SH}
\ee
and $\kappa$ is the surface gravity of the background black hole horizon. 
(iii) The angular momentum of the system measured at the black hole,
defined by 
\be
{\cal J}_H \equiv - \int_{\cal H}\bQ[\varphi].
\ee

In fact, the angular momentum can be measured either at 
the black hole horizon or at spatial infinity; 
since the metric is axisymmetric with axial killing field $\varphi^a$, 
it can be seen from (\ref{J}) that $\bJ[\varphi]$
vanishes, when pulled back to a slice to which $\varphi^a$ is tangent. 
This ensures (integrating the relation $\bJ[\varphi]=d\bQ[\varphi]$ 
over $\Sigma$) that, for the background solution,
\be
\int_{S^{\infty}}\bQ[\varphi] = \int_{\cal H}\bQ[\varphi].
\ee
In addition, by considering the identity 
\be
\int_{\Sigma} \om(\phi,\de\phi,{\cal L}_{\varphi}\phi)=
\int_{\dee\Sigma}\de\bQ[\varphi]-\varphi\cdot\syp,
\label{amid}
\ee
we see that when $\varphi^a$ Lie derives all
the dynamical fields, the left side of this equation vanishes.
Since $\varphi^a$ is tangent to the two-spheres $\cal H$
and $S^{\infty}$, the pullback of the second term 
on the right side vanishes. It follows that  
\be
\de \int_{S^{\infty}}\bQ[\varphi] = \de \int_{\cal H}\bQ[\varphi].
\label{Jsame}
\ee
Therefore, in spacetimes which have axisymmetric background configurations, 
the angular momentum measured at the black hole is equivalent to the
canonical angular momentum $\cal J$, measured at spatial infinity 
\be
{\cal J}\equiv-\int_{S^{\infty}}\bQ[\varphi],
\label{Jspi}
\ee
both when $\varphi$ is an axial killing field, (in the background solution) 
and for arbitrary solutions which are perturbations,
$\de \phi$, of the axisymmetric solution. This calculation also shows
that the definition of ${\cal J}_H$ is gauge independent, 
for arbitrary perturbations of an axisymmetric solution. This is because
$\de{\cal J}_H = 0$ when we choose $\de\phi$ to be pure gauge, 
which we see by first setting $\hat{\de}\phi\equiv\Liev\phi$ 
for some smooth $v^a$, and then 
replacing $\hat{\de}\phi$ with a gauge transform ${\hat{\de}}'\phi$
which coincides with $\hat{\de}\phi$ in a neighbourhood of the bifurcation
sphere, but vanishes in a neighbourhood of spatial infinity. 
Then we have for every $\hat{\de}\phi$ (using (\ref{Jsame})),
\bea
\hat{\de}{\cal J}_H = {\hat{\de}}\int_{\cal H}\bQ[\varphi]&=&
{\hat{\de}}'\int_{\cal H}\bQ[\varphi]\nonumber\\
&=&{\hat{\de}}' \int_{S^{\infty}}\bQ[\varphi] \nonumber\\
&=&0.
\eea
So we have that when $T^{ab}$ vanishes (along with $\bL_m$),
the interpretation of the terms in (\ref{prefl})
is straightforward. and one obtains a formula (\ref{genfl2})
which (bearing in mind the equivalence of $\cal J$ and ${\cal J}_H$)
is the formula (\ref{genfl}).

What if the set of fields $\psi$ is non-empty ?
In general, the ambiguity (\ref{Lambig}) in breaking $\bL$ into $\bL_g$ and 
$\bL_m$ stops us from meaningfully interpreting 
the surface terms in (\ref{prefl})
as perturbations of mass, entropy and angular momentum: even if
the overall theory is fixed, every choice of $\bL_g$ generates a 
different relation, with different choices of ${\bQ_g}$ etc. 
We therefore seek more restrictive assumptions under which we might
successfully identify the surface terms in (\ref{prefl}). 
One approach is to {\em fix} a particular choice 
of $\bL_g$ and think of it as specifying an independent 
theory. We assume there exists a form $\bB_g$ such that at spatial infinity,
$\de(\xi\cdot\bB_g)=\xi\cdot\syp_g$, and consider the functional $M_g$
defined by  
\be
M_g \equiv \int_{S_{\infty}}{\bQ_g}[\xi]-\xi\cdot\bB_g.
\ee
If we now require that the stress-energy of the matter 
distribution falls off sufficiently rapidly at spatial infinity, 
such that (near spatial infinity) the metric for any solution 
of the $\bL$-theory approaches  
a metric solution of the $\bL_g$-theory, and $M_g$ yields the same 
result on both metrics, then it makes sense to 
define the mass of the system as $M_g$.
We note that if we can also find a form $\bB(\phi)$ for the full 
theory, such that at spatial infinity $\de(\xi\cdot\bB)=\syp(\phi,\de\phi)$, 
then we can also define a canonical energy, $\cal E$, for the full 
theory given by (\ref{EH}), and in general ${\cal E}\ne M_g$. 

Therefore, when the stress-energy of the matter distribution falls 
off sufficiently rapidly, we can interpret the left side of (\ref{prefl}) 
as the variation of the mass of the system. The surface terms on the right side 
of (\ref{prefl}) are (variations of) the functionals that would
measure the entropy and angular momentum of a stationary black hole
in the $\bL_g$-theory. We might therefore be tempted to interpret
them as the black hole entropy and angular momentum; indeed, 
since ${\bQ_g}$ is the Noether charge of the $\bL_g$ theory,
we know from \cite{IW1} that one can define a quantity, $S_g$, by 
\be
S_g \equiv -2\pi\int_{\cal H} \frac{\de \bL_g}{\de R_{abcd}}\eps_{ab}\eps_{cd},
\ee
such that
\be
\de\int_{\cal H}{\bQ_g}[\chi]=\frac{\kappa}{2\pi}\de S_g.
\ee
One might also define a quantity, ${J_g}_H$, by
\be
{J_g}_H \equiv -\int_{\cal H}{\bQ_g}[\varphi].
\ee 
Although we made no assumptions about the axisymmetry of the matter fields,
we can show, providing the support of $T^{ab}$ does not intersect
some neighbourhood, $U$, of the bifurcation sphere, that ${J_g}_H$ is also 
well-defined (gauge independent) for arbitrary perturbations of the axisymmetric
solution. This follows by evaluating the left side of 
(\ref{amid}), using the fact that the 
calculation (\ref{omvol}) also holds when $\xi^a$ is replaced by $\varphi^a$.
Taking $\varphi^a$ to be tangent to the spatial slice, Eq. (\ref{amid}) 
then becomes
\be
\int_{S_{\infty}}\de{\bQ_g}[\varphi]+\int_{\Sigma}\de(\eps\cdot\; T\cdot 
\varphi)
=\int_{\cal H}\de{\bQ_g}[\varphi]
\label{Qphint}
\ee
Now, as before, let the perturbation in this equation be gauge, 
$\de\phi=\hat{\de}\phi$. Then we again can replace the perturbation on the
right side with an equivalent gauge change, which vanishes outside $U$, 
and so intersects neither the support of $T^{ab}$ nor
spatial infinity. Then we have the left side of (\ref{Qphint}) vanishes, and so
\be
\hat{\de}{J_g}_H=-\int_{\cal H}\hat{\de}{\bQ_g}[\varphi]=0.
\ee
Therefore ${J_g}_H$ is defined for arbitrary perturbations of an axisymmetric
solution.

Now having defined $M_g$, $S_g$ and ${J_g}_H$, we could write out 
(\ref{prefl}) in the form 
\be
\de M_g = 
\frac{\kappa}{2\pi}\de S_g +\Omega_H \de {J_g}_H
+ \int_{\Sigma}\frac{1}{2}\xi\cdot\eps\; T^{ab}\de g_{ab}
-\de(\eps\cdot\; T\cdot \xi), 
\label{myfl}
\ee
where $\de$ is a perturbation to an arbitrary nearby solution. 
However, we caution the reader that the identification of black 
hole entropy with
$S_g$ in general gives results in conflict with those in \cite{IW1}:
consider a theory of gravitation with a scalar field,
for which the matter Lagrangian couples to the spacetime curvature,
and which displays stationary black hole configurations in which the scalar 
field has sufficiently rapid spatial falloff.  We can therefore 
write out (\ref{myfl}) and interpret the black hole entropy as $S_g$.
From the results of \cite{IW1} we expect the entropy of the black hole
to {\em include} contributions from the scalar field; equation
(\ref{myfl}), however, defines a black hole entropy $S_g$ with only
metric contributions, with the entropy contribution of the scalar field 
somehow distributed in the volume integral of its stress-energy. These two
points of view are contradictory; therefore, while there are clearly
special cases (for instance, the Einstein-perfect fluid system) in which 
we can identify $S_g$ as the black hole entropy, and terms in the 
volume integral as (variations of) the matter entropy,
in general we regard
the notion of the black hole entropy defined by $S_g$ as inappropriate. 
Clarifying when $S_g$ can be correctly interpreted as black hole entropy 
is the subject of future research.

We note parenthetically that we can write out an alternative form 
of (\ref{myfl}) by replacing the stationary killing field $\xi^a$ in
(\ref{wfl}) with the horizon killing field $\chi^a$. (The analysis up
to (\ref{interfl}) is unchanged except for the substitution $\xi^a
\rightarrow\chi^a$.) Then expanding $\chi^a=\xi^a+\Omega_H\varphi^a$ at
spatial infinity and on the slice $\Sigma$, (but not at $\cal H$) 
and making the definitions discussed above gives the identity
\be
\de M_g = 
\frac{\kappa}{2\pi}\de S_g + \Omega_H \de {J_g}_{\infty}
+ \int_{\Sigma}\frac{1}{2}\xi\cdot\eps\; T^{ab}\de g_{ab}
-\de(\eps\cdot\; T\cdot \xi) 
-\Omega_H\int_{\Sigma}\de(\eps\cdot\; T\cdot \varphi) 
\label{myfl2}
\ee
where ${J_g}_{\infty}\equiv-\int_{S^{\infty}}\bQ_g[\varphi]$, is 
the system angular momentum measured at 
spatial infinity. Therefore the cost we
have incurred for the transfer of the angular momentum
integral to spatial infinity
is the appearance of an extra term in the volume integral. 

A relation of the form (\ref{myfl}), was first given by Schutz
and Sorkin \cite{ss}, in the case where $\bL_g$ was fixed to be
the Lagrangian for general relativity, $\bL_m$ was any matter 
Lagrangian, and there was no black hole boundary $\cal H$,
for the hypersurface $\Sigma$.
The relation stated in \cite{ss} is correct, but we comment here
on the ambiguity of the ``Noether operators" used by Schutz and Sorkin
to derive it: 
In its initial definition \cite{ss} the Noether operator for a Lagrangian
$\bL$ and a smooth vector field $\xi^a$ was defined to be
any (not necessarily covariant) three form $\bJ^S[\xi]$ satisfying the relation 
\be
\Liek \bL = \bE\Liek\phi + d(\bJ^S[\xi]+\xi\cdot\bL), 
\ee
for every smooth field vector field $\xi^a$.
This definition leaves $\bJ^S[\xi]$ ambiguous by an {\em arbitrary} exact 
three-form which is a linear differential operator in $\xi^a$. Since we know
from (\ref{eEx}) that $\bJ^S=d\bQ[\xi]$ when the field equations hold, 
this ambiguity would permit $\bJ^S=0$ 
as a valid Noether operator (which, following Schutz and Sorkin's 
methods, would yield a correct but trivial relation). On the
other hand, our definition of the Noether current admits a limited 
set of ambiguities (stated after (\ref{dJgen})), which 
cannot be used to annihilate the Noether charge, and in particular 
do not change the content of the first law.

Sorkin introduced an augmented definition of the Noether operator
in \cite{Sorkin92}, requiring that for a variation of the dynamical fields 
given by $\de\phi = f\Liek\phi$, where $f$ is any function, the
Noether operator $\bJ^{S'}$ be defined by
\be
\de\bL = \bE f\Liek\phi + d(f\bJ^{S'}[\xi]
+f\xi\cdot\bL). 
\ee
Providing one can find a $\bJ^{S'}$ which satisfies this relation, 
it is easy to see that one cannot add a term to $\bJ^{S'}$ which is
both exact and linear in $f$, for arbitrary $f$.
For a theory with a first order Lagrangian, finding such a $\bJ^{S'}$ is always 
possible: in \cite{Sorkin92} a first-order (noncovariant) Lagrangian for 
Einstein-Maxwell theory was used to yield an unambiguous 
Noether operator. It is not clear, however, that any general  
Lagrangian theory has a first order {\em Lagrangian} formulation,
so in general, Sorkin's definition may not even yield a Noether operator.
In contrast, all of our Noether currents $\bJ[\xi]$
defined above can be computed for Lagrangian theories of arbitrary
derivative order, and are manifestly covariant, 
requiring no additional background 
fields (apart from the symmetry field $\xi^a$) for their definition. 
For these reasons, we feel that whilst our relation (\ref{myfl})
and that in \cite{ss} coincide for an Einstein-matter system
without the black hole, (\ref{myfl}) is defined more generally.

We finally remark that we could have carried out the entire
analysis leading up to (\ref{prefl}) allowing the Lagrangian 
$\bL_g$ to depend on a {\em set} of 
stationary axisymmetric fields, {$s_i$}, including the metric, 
and the Lagrangian $\bL_m$
to depend on $s_i$ and a distinct set of fields, $\psi$, which didn't appear 
in $\bL_g$, to obtain a relation very similar to (\ref{prefl}). 
The resulting perturbative identity 
has the terms ${\bQ_g}$ and $\syp_g$ in (\ref{prefl}) replaced with 
the Noether charge and symplectic potential in the theory described by
$\bL_g$ (which now depends on both the metric and the other matter 
fields in the set {$s_i$}), and the volume term is now given by
\be
\int_{\Sigma}\frac{1}{2}\xi\cdot\eps\;T_{s_i}\de s_i
-\de(\eps\cdot\;T_s\cdot s\cdot \xi)
\ee
where the first term in the volume integral is defined by 
the variation of $\bL_m$:
\be
\de\bL_m=\bE_m\de \psi+ \frac{1}{2}T_{s_i}\de s_i + d\syp_m(\phi,\de\phi).
\ee
Giving each $s_i$ field $u_i$ upper, and $d_i$ lower indices, in the manner
\be
s_i\rightarrow{{s_i}^{b_{u_1}\cdots b_{u_i}}}_{a_1\cdots a_{d_i}},
\ee
the second term in the volume integral is defined by 
\bea
(\eps\cdot T_s\cdot s\cdot\xi)_{abc}&\equiv&
\eps_{eabc}\sum_i[ 
{{T_{s_i}}_{b_1 \cdots {b_u}_i}}^{a_1 \cdots {a_d}_i} (
-{{s_i}^{e \cdots {b_u}_i}}_{a_1 \cdots {a_d}_i}\de^{b_1}_p
\ldots -{{s_i}^{b_1 \cdots e}}_{a_1 \cdots {a_d}_i}\de^{b_{u_i}}_p
\nonumber\\
&&+{{s_i}^{b_1\cdots {b_u}_i}}_{p \cdots {a_d}_i}\de^{e}_{a_1}
\ldots +{{s_i}^{b_1 \cdots b_{u_i}}}_{a_1 \cdots p}\de^e_{a_{d_i}})\xi^p].
\nonumber\\
\label{eTx}
\eea

\section{A review of perfect fluids, and three variational formulations.}
In this section we recall the definition, the relevant properties, 
and three variational principles for a self-gravitating perfect fluid: 
one given by Schutz \cite{SchF}, (which we use 
in the Appendix to derive
a conserved current for perturbations of Einstein-perfect fluid systems), 
the ``axionic vorticity" formulation given by Carter \cite{carter} 
for an isentropic perfect fluid
(which we use in the next section, to derive a first law), and 
a ``convective" approach also described by Carter \cite{carter}. 
Our aim is to gather the results we need for the calculations of 
the following sections; detailed treatments of these variational
principles can be found in \cite{SchF,carter,brown}.
 
From the viewpoint of black hole mechanics, we would like a stationary
axisymmetric black hole configuration to be represented by a Lagrangian
theory in which all the fields appearing in the Lagrangian (the 
{\em dynamical} fields) are also stationary and axisymmetric.
Having stated these formulations, however, we will see that they 
all have fluid configurations in which the {\em physical} fields 
(the fluid four velocity, number density, entropy and functions 
of these fields) are stationary and axisymmetric, but in which   
the dynamical fields possibly share neither of these symmetries. 
The question as to whether a variational 
principle exists that always represents 
(physically) stationary axisymmetric configurations with dynamical fields
that also have these properties is (as far as we are aware) open.

By a perfect fluid on a fixed spacetime background \cite{brown,MTW}
we mean a system described by five scalar fields, $(n,s,\rho,p,T)$,
on spacetime and one (unit, timelike) vector field $U^a$, such that
$\rho=\rho(n,s)$ is a fixed function, and the following equations 
hold on the fields: the first law of thermodynamics,
\be
d\rho(n,s)=\frac{(p+\rho)}{n} dn + nT ds,
\label{flrhons}
\ee
and the equations of motion,
\be 
\del_a(nU^a) = 0 \mbox{,\qquad and \qquad} \del_aT^{ab}=0,
\label{Tfe}
\ee
where $T^{ab}$ is defined by 
\be
T^{ab} \equiv (p+\rho) U^aU^b + p g^{ab}.
\label{Tab}
\ee
The fields $n,\rho,s,p,T$ and $U^a$ 
have physical interpretations as the number density, 
energy density, entropy per particle (specific entropy), 
pressure, temperature, and four velocity of the fluid, respectively.

We note that (\ref{Tfe}) can be given a useful alternative form,
by first defining the specific inertial mass $\mu$: 
\be
\mu \equiv \frac{p+\rho}{n} 
\ee
which along with (\ref{flrhons}) implies
\be
d p = n d\mu - nTd s.
\label{flpmus}
\ee
By using these relations in the second equation of (\ref{Tfe}) we get
(see \cite{brown}) an equivalent pair of equations of motion - 
\be
\del_a(nU^a) = 0 \mbox{,\qquad and \qquad} nU^a\omega_{ab}= nT\del_b s,
\label{vortfe}
\ee
where the fluid vorticity two-form, $\omega_{ab}$, is defined as
\be
\omega_{ab}\equiv 2\del_{[a}(\mu U_{b]}).
\label{vortdef}
\ee

If desired, one can define the entropy per unit volume, $S$
(entropy density), by $S=n s$. Substituting this definition of $S$ into
(\ref{flrhons}) 
and defining the chemical potential, $\mu'$, by
\be
\mu'\equiv \frac{p+\rho - TS}{n},
\label{mupdef}
\ee
then gives the relation
\be
d\rho (n,S)=\mu'dn + TdS.
\label{flrhonS}
\ee

We now specify three variational formulations for this perfect
fluid, over a fixed spacetime background 
(coupling the theories to gravitation amounts to
adding the appropriate metric Lagrangian, which we do later). 
First, we state the ``velocity-potential" representation
of Schutz \cite{SchF}: here the dynamical fields
of the fluid are given by scalars $\Phi,\alpha,\beta,\theta$,and $\sigma$.
One now defines a function $m$ which depends on these fields via
the relation 
\be
m^2 = -(\del_a\Phi + \alpha\del_a\beta + \theta\del_a \sigma)
(\del^a\Phi + \alpha\del^a\beta + \theta\del^a \sigma),
\ee
and the fluid Lagrangian is given by
\be
\bL_f\equiv\eps P(m,\sigma).
\label{Lepf}
\ee
where $P(m,\sigma)$ is some fixed function.
One can verify \cite{SchF,brown} that we recover (\ref{flrhons}), and 
also that 
the equations of motion for the fields $\Phi,\alpha,\beta,\theta,\sigma$
arising from this Lagrangian reduce to (\ref{vortfe}), provided one
defines the physical fields in terms of the dynamical fields in these
equations by: 
\bea
P&\rightarrow&p\nonumber\\
m&\rightarrow&\mu\nonumber\\
\sigma&\rightarrow&s\nonumber\\
(\dee P/\dee m)_{\sigma}&\rightarrow&n\nonumber\\
(\dee P/\dee \sigma)_{m}&\rightarrow&-nT\nonumber\\
\del_a\Phi+\alpha\del_a\beta+\theta\del_a\sigma&\rightarrow&\mu U_a.
\label{Udef}
\eea
Conversely, given any configuration of the 
physical fields $(n,\rho,s,p,T,U^a)$ satisfying
(\ref{flrhons}) and (\ref{Tfe}), it can be shown (see \cite{SchF}) that 
there exist functions $(P,m)$ and (non-unique)
dynamical fields $(\Phi,\alpha,\beta,\theta,\sigma)$ related
to the physical fields by (\ref{Udef}), which satisfy the equations
of motion arising from Lagrangian (\ref{Lepf}).

Next, Carter's variational formulation \cite{carter} for an {\em isentropic}
perfect fluid, (by which we mean that the fluid has an everywhere 
constant specific entropy $s$), 
defines the dynamical fields to be a two-form and two scalars, 
$b_{ab}$ and $\chi^{\pm}$. 
The fluid Lagrangian is given in terms of these fields by 
\be
\bL_f=[-r(\nu)-\frac{1}{2}\eps^{abcd}b_{ab}\del_{c}\chi^+\del_{d}\chi^-]\eps.
\label{Lm}
\ee
where the function $r(\nu)$ is fixed, 
and the function $\nu$ is defined in term of the potentials $b_{ab}$
by the relation
\be
\nu^2 \equiv \frac{3}{2}(\del_{[a}b_{bc]})(\del^{[a}b^{bc]}).
\ee
As shown in \cite{carter}, 
if one defines the physical fields as follows:
\bea
r&\rightarrow&\rho\nonumber\\
\nu&\rightarrow&n\nonumber\\
\nu(\dee r/\dee \nu)- r&\rightarrow&p,\nonumber\\
3 \del_{[c} b_{ab]}&\rightarrow&N_{abc},
\label{b}
\eea
where the number-density three-form $N_{abc}$ is given by  
\be 
N_{abc} = n \eps_{abcd}U^d,
\label{N}
\ee
then we recover (\ref{flrhons}), and
the field equations for $b_{ab}$ and $\chi^{\pm}$
yield the second equation in (\ref{vortfe}) in the case $\del_a s=0$, 
as well as the relation 
\be
\omega_{ab}=2 \del_{[a}\chi^+\del_{b]}\chi^-.
\label{vortchi}
\ee
Given relation (\ref{b}) between $b_{ab}$ and $N_{abc}$,
one sees that the first equation in (\ref{vortfe}) is satisfied vacuously,
since it can be rewritten as
\be
\del_{[a}N_{bcd]}=0,
\label{eqs}
\ee
but the definition of $N_{abc}$ shows $d{\bf N}=dd{\bf b}=0$ automatically.

A third type of variational formulation given by Carter \cite{carter},
and treated in more detail by Brown \cite{brown},
(which is the equivalent diffeomorphism invariant version of the formalisms
specified by Taub \cite{Taub}, or Hawking and Ellis \cite{HE}),
has dynamical fields $X^{A}$ for $A = 1,2,3$. In this 
formalism one must specify two functions- $r(\nu,\sigma)$, and $\sigma(X)$, 
where $\nu$ is defined in terms of the $X^A$ by 
\be
\nu^2 \equiv 6 
[N_{ABC}(X)\del_aX^A\del_bX^B\del_cX^C][N_{DEF}(X)\del^aX^D\del^bX^E\del^cX^F],
\ee
and $N_{ABC}(X)$ is a fixed three-form on the three-dimensional
manifold which has $X^A$ as coordinate fields. 
The Lagrangian is then given by 
\be
\bL_f = - \eps r(\nu,\sigma).
\ee
The equations resulting from this Lagrangian for the fields $X^A$ are
seen to reduce to the second equation in (\ref{vortfe}) after one 
has set 
\bea
r&\rightarrow&\rho\nonumber\\
\nu&\rightarrow&n\nonumber\\
\sigma&\rightarrow&s\nonumber\\
\nu(\dee r/\dee \nu)_{\sigma}- r&\rightarrow&p\nonumber\\
(\dee r/\dee \sigma)_{m}&\rightarrow&nT\nonumber\\
N_{ABC}(X)\del_aX^A\del_bX^B\del_cX^C&\rightarrow&N_{abc},
\eea
where $N_{abc}$ is defined from (\ref{N}).
(This relation between the physical $N_{abc}$ and the dynamical
fields also ensures that $N_{abc}$  is automatically conserved.) 
The $X^A$ are interpreted as coordinates on a 
``base manifold", obtained by treating the spacetime as a bundle with fibres
given by the integral curves of the four-velocity. We will not use 
this formulation for two reasons: firstly, the assignment of the entropy, $s$,
as a fixed function of the $X^A$ only allows us to perturb it by 
diffeomorphisms of the base manifold
(for this reason we use Schutz's formalism for the calculation 
in the Appendix). Secondly, it is unclear that there are {\em any}
solutions in which the $X^A$ are globally well-defined axisymmetric fields
on spacetime (for this reason, in section 4, we use the 
formulation due to Carter with Lagrangian (\ref{Lm})).

In order to write the first law in form (\ref{genfl}), only involving
surface integrals, we must assume that all the dynamical fields are
stationary and axisymmetric in the background solution. Now even if 
a fluid configuration has stationary and axisymmetric {\em physical}
fields (the fluid number density, entropy and functions of these fields), 
the {\em dynamical} fields (the fields appearing in the Lagrangian)
corresponding to these physical fields
may not possess these symmetries. Therefore, the requirement 
of stationarity and axisymmetry on the dynamical
fields may restrict the choice of background configurations.
In fact, for Schutz's formulation, we see from 
the definition of the four velocity (\ref{Udef}) that
physical fluid configurations with an everywhere causal four-velocity
(including those which are stationary and axisymmetric) must 
include at least one nonstationary dynamical field. There are therefore
no physically interesting fluid configurations in which all the dynamical
fields in this formulation are stationary.

On the other hand, for Carter's formulation, 
it is evident that there must be {\em some} 
physically stationary fluid configurations with stationary dynamical fields; 
(for instance, a static spherically symmetric fluid distribution 
could have the field $b_{ab}$ given by
${\bf b}\sim f(r) {}^2\eps$ and $\chi^{\pm}=0$,
where ${}^2\eps$ is the volume element on the spheres of symmetry). 
However, we will see in the next section 
(in the discussion above (\ref{circf})) that a stationary, 
axisymmetric, circular flow (in a spacetime which also has these symmetries) 
must be vortex-free, if $\chi^{\pm}$ are restricted to be
stationary and axisymmetric.
That is, the assumption of 
stationarity and axisymmetry on the vorticity potentials $\chi^{\pm}$ 
restricts the allowed stationary axisymmetric 
configurations a fluid can adopt. We make no attempt here 
to enumerate the set of physically stationary 
and axisymmetric configurations which also have these symmetries in the 
dynamical fields (or indeed, in the case of black hole spacetimes, to
investigate whether this set is non-empty). 
Rather, in the following section we will assume the potentials are 
stationary and axisymmetric, and write out the resulting first law
involving only surface terms, looking for any non-trivial 
modifications arising from the fluid fields.

We are unaware of a variational formulation for a 
perfect fluid which represents all stationary axisymmetric fluid 
configurations with stationary axisymmetric dynamical fields.
If it exists, then the following argument by Schutz and Sorkin \cite{ss}
shows that certain
compactly supported perturbations of the physical fields must correspond
to non-compactly supported perturbations of the dynamical fields. 
Since the calculation given in (\ref{omvol}) does not depend on 
the fulfillment of the field equations for $g_{ab}$, it is still valid
if we consider the fields $\psi$ to be the dynamical fields for a perfect
fluid over a fixed spacetime background, and we let $\de\psi$ be a 
perturbation to a nearby solution of the perfect fluid equations, with
$\de g_{ab}=0$. Now consider a formulation for a perfect fluid where,
for a general configuration in which all the physical fields (and 
the metric of the spacetime background) are stationary,
all the dynamical fields are also stationary. 
Then the left side of (\ref{omvol}) vanishes, and integrating the 
right side over a spatial slice $\Sigma$, we are left with
\be
\int_{\Sigma}\de(\eps\cdot T\cdot\xi) = 
\int_{\dee\Sigma}\de\bQ_m[\xi]-\xi\cdot\syp_m(\phi,\de\psi).
\label{prenop}
\ee
This implies that for perturbations of the physical fields for which 
the corresponding perturbations of the dynamical fields are {\em compact},
we must have 
\be
\int_{\Sigma}\de(\eps\cdot T\cdot\xi) = 0,
\label{nop}
\ee
which, for a perfect fluid, is clearly false
for a general stationary background. This implies
that if a variational formulation is to have dynamical fields which are
always stationary when the physical fields are stationary, then perturbations
of the physical fields which yield a non-zero result on the left 
side of (\ref{prenop}) must correspond to spatially non-compact
perturbations of the dynamical fields. This requirement rules out the
existence of a variational principle in which the physical fields 
{\em are} the dynamical fields \cite{ss}. However, the existence
of a variational principle for a perfect fluid in which all 
configurations with stationary and axisymmetric physical fields are
represented by dynamical fields with these symmetries is still an open
question.

\section{First laws of black hole mechanics with perfect fluids}
We now present two forms of the first law of black hole mechanics
which incorporate perfect fluids. The first form is a special case of 
the perturbative identity (\ref{myfl}), where $\bL_g$ 
is the usual Hilbert Lagrangian for general relativity, 
and $\bL_m$ is any Lagrangian for a perfect fluid. 
This form of the first law allows non-stationary dynamical fields, at the cost 
of having volume integrals in the interior of the spacetime. We then
compute a second form of the first law only involving surface integrals 
for both metric and fluid fields, using Carter's variational 
formulation presented above, and the methods of \cite{IW1}.

\subsection{The first law with volume integrals}
We now write out the perturbative relation (\ref{myfl}), 
setting $\bL_g =1/16\pi R$, and $\bL_m$ to be any perfect fluid Lagrangian 
which allows all possible perturbations of the physical fields of
the perfect fluid off an arbitrary background.
(From the comments below Eq.(\ref{Udef}) it is 
evident that Schutz's variational formulation, with Lagrangian (\ref{Lepf})
satisfies this criterion.) As stated in Lemma 2, 
we assume the metric of the background spacetime is asymptotically flat, 
stationary and axisymmetric with a stationary killing field
$\xi^a$ and axial killing field $\varphi^a$. We also assume the existence of
a bifurcate killing horizon, 
with horizon killing field $\chi^a=\xi^a+\Omega_H\varphi^a$, where $\Omega_H$
is the angular velocity of the horizon.

In this case (see \cite{IW1}) the term
$M_g$ in (\ref{myfl}) can be shown to be the ADM mass, $S_g$
to be $1/4 A_H$, and ${J_g}_H$ the expression $J_H$ for black 
hole angular momentum given in (\ref{bchfl}).
The terms involving the stress-energy tensor 
have been shown by Bardeen, Carter and Hawking \cite{bch} 
to reduce to the fluid terms in (\ref{bchfl}), 
but for completeness (and to fix the signs for our choice of orientations)
we briefly demonstrate this fact: 
in \cite{bch} the four velocity of the fluid 
with angular velocity $\Omega$ (which need not be constant)
was set to be $U^a = v^a/(-v\cdot v)^{1/2}$ where $v^a=\xi^a+\Omega\varphi^a$. 
Now using (\ref{Tab}),(\ref{mupdef}) and (\ref{flrhonS}), (assuming, as
usual, that we identify the perturbed spacetime such that 
$\de\xi^a = \de\phi^a = 0$), one obtains 
\bea
\de({T^a}_b\xi^b\eps_{apqr})&=&v^b\de{T^a}_b\eps_{apqr}
-\Omega\de ({T^a}_b\varphi^b\eps_{apqr})\nonumber\\
&=&v^a \de ((\mu' n + TS)v_a(-v\cdot v)^{-1/2}U^b\eps_{bpqr}+p\eps_{apqr})
-\Omega\de ({T^a}_b\varphi^b\eps_{apqr})\nonumber\\
&=&(p+\rho)v^a\de(v_a(-v\cdot v)^{-1/2})U^b\eps_{bpqr} + \frac{1}{2}
p g^{cd}\de g_{cd}\xi^a\eps_{apqr} \nonumber\\
&&-\mu'(-v\cdot v)^{1/2}\de(nU^b\eps_{bpqr})
-T(-v\cdot v)^{1/2}\de(SU^b\eps_{bpqr})\nonumber\\
&&-(n\de\mu' + S\de T) \xi^a\eps_{apqr}
+v^a\eps_{apqr}\de p- \Omega\de ({T^a}_b\varphi^b\eps_{apqr})\nonumber\\
&=& \xi^a\eps_{apqr}\frac{1}{2}T^{cd}\de g_{cd} 
+\mu'(-v\cdot v)^{1/2}\de(nU^b\eps_{pqrb})
-T(-v\cdot v)^{1/2}\de(SU^b\eps_{bpqr})\nonumber\\
&& -\Omega\de ({T^a}_b\varphi^b\eps_{apqr}),
\label{bchcalc}
\eea
When all these substitutions are inserted into (\ref{myfl}), it 
reduces to 
\be
\de M = \frac{\kappa}{8\pi}\de A + \Om_H \de J_H - \int_{\Sigma}\mu'|v|
\de N_{abc} + \int_{\Sigma}\Om\de J_{abc}+\int_{\Sigma}T|v|\de S_{abc},
\label{final}
\ee
which is identical to (\ref{bchfl}), except that $\de$ now
represents an arbitrary perturbation (not necessarily stationary
or axisymmetric) of the background. In this sense, (\ref{final})
is a generalisation of (\ref{bchfl}).

\subsection{A (restricted) first law with surface integrals}
In the previous section we observed that the variational formulations 
we presented were constrained in the stationary axisymmetric
fluid configurations they could represent, given the requirement that 
their dynamical fields obeyed these symmetries. One might therefore 
suspect that any form of the first law involving only surface integrals  
could not include non-trivial fluid contributions.
Indeed, if we add Schutz's Lagrangian 
(\ref{Lepf}) to the Lagrangian of an arbitrary metric theory of gravity,
and construct a first law using the analysis of \cite{IW1} then 
we find {\em no} additional contributions to this first law 
from the fluid fields, providing the fluid's number density decays sufficiently
rapidly at spatial infinity, and does not intersect the black hole horizon. 
It is possible, however, to convert {\em some} of the volume
integrals in (\ref{bchfl}) into surface integrals, by choosing Carter's
variational formulation (\ref{Lm}). We do so below, finding a first
law for an arbitrary metric theory of gravity coupled to an isentropic
perfect fluid, in which the background configuration for the perfect fluid
as well as the allowed perturbations of the physical fields 
are restricted. (Note that the gravitational contributions to 
such a first law have been considered in detail in \cite{IW1}. 
We are interested in the fluid contributions.) We finally verify that this
first law reduces to (\ref{bchfl}) when the assumptions made in the two
derivations overlap. Our first law is the following result:

{\bf Lemma 3:} 
Let $\bL$, given by
\be
\bL = \bL_g - \eps(r(\nu)+
\frac{1}{2}\eps^{abcd}b_{ab} \del_{[c}\chi^+\del_{d]}\chi^-),
\label{L}
\ee
be the Lagrangian for an isentropic perfect fluid coupled to an arbitrary metric
theory of gravity, where 
$\bL_g=\eps L_g(g_{ab},R_{abcd},\del R_{abcd},\ldots,(\del^p)R_{abcd})$,
and the perfect fluid formulation, with dynamical fields 
$(b_{ab},\chi^{\pm})$, is summarised below (\ref{Lm}).
Fix an asymptotically flat black hole solution with bifurcate killing
horizon, with the spacetime structure and the killing fields 
described in Lemma (2), with the 
additional assumptions that {\em all} the dynamical fields, (not just
the metric) in this theory are stationary and axisymmetric, and that
all the dynamical fields are globally defined.
Let $\de\phi$ be a perturbation of the dynamical fields, from such
a solution to an arbitrary nearby solution, with $\de\xi^a=0$. 
With these assumptions
the following identity is the first law of black hole mechanics
for this system: 
\be
\de M_g = \frac{\kappa}{2\pi}\de{\cal S} 
+\Omega_H \de {\cal J}_H
+\int_{\cal H} \mu_{\infty} \de b_{qr},
-\int_{S^{\infty}} \mu_{\infty} \de b_{qr},
+\int_{\cal H} X_{qr} - \int_{S^{\infty}} X_{qr}
\label{fl}
\ee
Here we define the mass of the system, $M_g$, as 
\bea
M_g&\equiv& \int_{S_{\infty}}{\bQ_g}[\xi^a]-\xi\cdot\bB_g 
\label{m}
\eea
and the entropy, $\cal S$, and angular momentum, ${\cal J}_H$, 
of the system by  
\bea
{\cal S}&\equiv&-2\pi\int_{\cal H} \frac{\de\bL_g}{\de R_{abcd}}\eps_{ab}
\eps_{cd}\nonumber\\
{\cal J}_H&\equiv&-\int_{\cal H} {\bQ_g}[\varphi],
\eea
where $\kappa$ is the surface gravity of the black hole,
the two-form $\bQ_g[\xi]$ was defined in (\ref{Jstruct}),
and the three-form $\bB_g$ is such that, at spatial infinity, 
$\de(\xi\cdot\bB_g) =\xi\cdot\syp_g$, with $\syp_g$ given by (\ref{deLm}).
Finally, the two-form $X_{qr}$ is defined by 
\be
X_{qr}\equiv 2\xi^p b_{p[q}[\de(\mu U_{r]}) -
\del_{r]}\chi^-\de\chi^++\del_{r]}\chi^+ \de\chi^-],
\label{X}
\ee
{\bf Proof:}\\
The first law of black hole mechanics in \cite{IW1}
is essentially given by the right side of (\ref{wfl}), 
when the left side vanishes because
of the assumed symmetries of the background fields. 
We therefore compute the quantities appearing in the right side 
of (\ref{wfl}): 
Varying the dynamical fields in $\bL$ (and performing the substitutions 
(\ref{b}) where applicable) yields the equations of motion and
the symplectic potential $\syp$:
\bea
\de \bL&=&\eps\left(\frac{\de L_g}{\de g_{ab}}+\frac{1}{2}T^{ab}\right)
\de g_{ab}+ 
\frac{1}{6}\eps N_{abc}\eps^{abcd}(\del_d\chi^+\de\chi^-
-\del_d\chi^-\de\chi^+)\nonumber \\
&&+\eps[\del_c(\frac{\mu}{2n}N^{abc})
-\frac{1}{4}\eps^{abcd}\omega_{cd}] \de b_{ab} + d\syp,
\label{delL}
\eea
with the stress-energy tensor
\be
T^{ab} = \frac{\mu}{2n}{N^a}_{cd}N^{bcd} - \rho g^{ab},
\label{tcart}
\ee
and the symplectic potential
\be
\syp_{pqr}(\phi,\de\phi)={\syp_g}_{pqr}(g,\de g) 
-\frac{\mu}{2n}N^{abc}\de b_{bc} \eps_{apqr}
+\frac{1}{2}\eps_{apqr} b_{cd}\eps^{abcd}(\del_b\chi^+\de\chi^-
-\del_b\chi^-\de\chi^+).
\label{theta}
\ee
It can be verified that the equations of motion for the fluid fields 
reduce to (\ref{vortfe}) using the definitions (\ref{b},\ref{N}). 
The stress-energy tensor (\ref{tcart}) is also seen to reduce to the usual 
form (\ref{Tab}) by expanding its first term:
$$\frac{\mu}{2n}{N^a}_{cd}N^{bcd}=\frac{\mu}{2n}nU^e{\eps^a}_{cde}
\eps^{bcdf}nU_f =  \mu n(g^{ab}+U^a U^b).$$
The Noether current associated to $\xi^a$ is
\be
\bJ_{pqr}[\xi] = {\bJ_g}_{pqr}[\xi] -(\frac{\mu}{2n}N^{dbc}N_{ebc}\xi^e
-\rho \xi^d)\eps_{dpqr} - \del_b(\frac{\mu}{n}N^{dbc}\xi^e
b_{ec}\eps_{dpqr}).
\label{N1}
\ee
Therefore, the integrand on the right side of (\ref{wfl}) evaluates to
\bea
(\de \bQ[\xi] - \xi\cdot\syp)_{qr} &=&
\de{\bQ_g}_{qr}[\xi] - \de(\frac{\mu}{2n}N^{abc}b_{ec}\xi^e\eps_{abqr})
-\xi^p[{\syp_g}_{pqr} - \frac{\mu}{2n}N^{abc}\de b_{bc} \eps_{apqr}\nonumber\\
&&
+\eps_{apqr}\frac{1}{2}\eps^{abcd}b_{cd}(\del_b\chi^-\de\chi^+-\del_b\chi^+
\de\chi^-)]\nonumber\\
&=&
\de{\bQ_g}_{qr}[\xi] -\xi^p{\syp_g}_{pqr}-\xi^p U_p\mu \de b_{qr} \nonumber\\
&&- \frac{1}{2}b_{qr}\xi^p(\del_p\chi^-\de\chi^+-\del_p\chi^+ \de\chi^-)+X_{qr}
\label{bterm}
\eea
where we define the two-form $X_{qr}$ by (\ref{X}),
and we used the identification $N_{abc}\equiv \eps_{abcd}nU^d$ to obtain 
the second line of (\ref{bterm}). 

When the background solution is a black hole with the structure
and symmetries specified in the statement of the Lemma, 
the fourth term in the second equation of
(\ref{bterm}) vanishes because the dynamical
fields are stationary: $\xi\cdot\del\chi^{\pm}=0$.
Now given the definition of vorticity (\ref{vortdef}) and its 
relation to the potentials
(\ref{vortchi}), it is evident that (locally) there exists some function $f$
such that $\mu U_a$ can be rewritten 
\be
\mu U_a = \del_a f+\chi^+ \del_a\chi^-.
\label{muU}
\ee
Let $t$ be a function such that $\xi^adt_a=1$.
Then the requirements that the four-velocity be causal, stationary and
axisymmetric, along with the assumed stationarity and axisymmetry
of $\chi^{\pm}$ force $f$ to be a sum of terms, 
one of which is strictly linear in $t$ (we define the constant of 
proportionality to be $-\mu_{\infty}$). 
For the same reason the $\varphi$-dependence 
of $f$ must be also linear, but this dependence can be ruled
out because the occurrence of such a term would force $U^a$ to be acausal
near spatial infinity. We therefore have that the form of $f$ is 
\be
f = -\mu_{\infty}t + g.
\label{fform}
\ee
where $\xi\cdot\del g = \varphi\cdot\del g = 0$. Therefore we see that 
the assumption of stationarity and axisymmetry on the dynamical 
fields (taking the four-velocity to be everywhere causal) has 
restricted us to a very narrow range of allowed background four-velocities;
for instance, we must have $\varphi^aU_a=0$. 
Moreover, when the vacuum theory is general relativity, 
with the flow assumed to be {\em circular} 
(tangent to the $\xi-\varphi$ subspaces),
there is only one possible solution: for this theory the subspaces orthogonal
to $\xi^a$ and $\varphi^a$ are integrable, 
and the resulting submanifolds can be endowed with 
coordinates $(x^1,x^2)$, such that the metric is
``block diagonal" with no ``cross-terms" between the subspace 
spanned by $\xi^a,\varphi^a$ and its orthogonal complement (see
Chapter 7 of \cite{WaldGR}). Now  
the assumption of circular flow forces $g=0$ and $\chi^+d\chi^-=0$, leaving
us with only
\be
\mu U_a = -\mu_{\infty} dt_a.
\label{circf}
\ee

In any case, using just the form of $f$ in (\ref{fform}), we see 
\be
\xi^a\mu U_a = -\xi^a\mu_{\infty}dt_a = -\mu_{\infty},
\ee
and the boundary term (\ref{bterm}) reduces to 
\be
\de \bQ_{qr}[\xi] - \xi^p\syp_{pqr} =  
\de {\bQ_g}_{qr}[\xi] - \xi^p{\syp_g}_{pqr} + \mu_{\infty} \de b_{qr} +X_{qr},
\label{btfinal}
\ee
We now assume the existence of a form $\bB_g$ such that at
spatial infinity $\xi\cdot{\syp_g}=
\de(\xi\cdot \bB_g)$, and write out the first law of black hole mechanics 
by substituting (\ref{btfinal}) into the surface integrals on the right side of
(\ref{wfl}), observing that the left side of (\ref{wfl}) vanishes due to
the symmetries assumed on the dynamical fields.
If we expand $\xi^a=\chi^a-\Omega_H\varphi^a$ at the bifurcation sphere
for the first two terms of (\ref{btfinal}), 
then we obtain (\ref{fl}) which is what we wished to show. $\Box$

The results of \cite{IW1} predicted that the first law (\ref{fl})
would only contain
surface integrals, and we see this is indeed the case. Note, however, that
the assumptions made about the symmetry of the dynamical fields 
restricted the allowed background fluid configurations for the fluid fields.
Moreover, by perturbing the local form 
of $\mu U_a$ in (\ref{muU}) we see that the
restriction to stationary and axisymmetric $\chi^{\pm}$ in background 
also prevents us from achieving all possible
perturbations of $\mu U_a$, by perturbing only the dynamical fields 
$b_{ab}$ and $\chi^{\pm}$. Finally both the background and the perturbed
configurations must be restricted such that the integral
$\int_{S^{\infty}} X_{qr}$ converges. (This, along with the following
result relating this term to the fluid angular momentum will guarantee 
the convergence of the corresponding boundary term at the bifurcation
sphere).

We finally show that (\ref{fl}) reduces to (\ref{bchfl}) when the
assumptions made in the two derivations overlap.
From our discussion in the last section we 
know that $M_g, {\cal S}$ and ${\cal J}_H$
reduce to their values for general relativity given in (\ref{bchfl}), when
$L_g = (1/16\pi) R$. We start by considering the fluid contribution in our  
first law (\ref{fl}) from the integral  
\bea
\de\int_{\infty} \mu_{\infty} b_{qr}-\de\int_{\cal H}\mu_{\infty} b_{qr}
&=&\de\int_{\Sigma} \mu_{\infty} N_{pqr}\nonumber\\
&=&\int_{\Sigma} \mu |v| \de N_{pqr},
\eea
where the last line follows because the fluid flow in \cite{bch} is assumed
to be tangent to the subspaces spanned by $\xi^a$ and $\varphi^a$: 
so taking the velocity to be $U^a = v^a/|v|$ where
$v^a = \xi^a + \Omega \varphi^a$, we see from the discussion above (\ref{circf})
that $\mu = -\mu U\cdot U = \mu_{\infty}v\cdot dt/|v| 
= \mu_{\infty}/|v|$, and so $\mu |v| = \mu_{\infty}$. 
Our first law now takes the form 
\be
\de M = \frac{\kappa}{8\pi}\de A+\Omega_H\de J_H
-\int_{\Sigma}\mu|v|\de N_{abc} 
+\int_{\cal H} X_{qr} - \int_{S^{\infty}} X_{qr}
\label{tmp1}
\ee

We now concentrate on the original form of the first law in
(\ref{bchfl}) and show that it agrees with (\ref{tmp1}). 
By repeating the calculation 
(\ref{bchcalc}) using the relation (\ref{flrhons})
instead of (\ref{flrhonS}) along with the assumption $\de s=0$
(as befits an isentropic fluid), we find the form 
of (\ref{bchfl}) for an isentropic fluid - 
\be
\de M = \frac{\kappa}{8\pi}\de A + \Om_H \de J_H - \int_{\Sigma}\mu|v|
\de N_{abc} +\int_{\Sigma}\Omega\de J_{abc}.
\label{tmp2}
\ee
Next, we demonstrate that the pullback to $\Sigma$ of the 
angular momentum density given 
in (\ref{tmp2}) reduces to the exterior derivative of the two 
form $X_{qr}$ defined in (\ref{X}), given the assumption that 
the dynamical fields are stationary and axisymmetric, i.e.,
\be
\Omega \de J_{pqr}=-(dX)_{pqr},
\ee 
where both sides are assumed pulled back to $\Sigma$.
To do this we compute the exterior derivative of (\ref{X}), finding
\be
(dX)_{pqr}= 3\xi^e N_{e[pq}(\de(\mu U_{r]}) -
\del_{r]}\chi^-\de\chi^++\del_{r]}\chi^+ \de\chi^-),
\label{dX}
\ee
where we have assumed $\Liek b_{ab} =0$.
Pulling this form back to $\Sigma$ by contracting with 
$1/6\eps^{spqr}n_s$ (where $n_s$ is the unit normal to $\Sigma$) yields
\be
\overline{dX} = -{}^3\eps (2 n n_e \de(\mu U_r) U^{[e}\xi^{r]}
+2\xi^{[e}U^{r]}n_e(\del_r \chi^- \de \chi^+ - \del_r\chi^+\de \chi^-)),
\ee
where ${}^3\eps$ is the volume form induced on $\Sigma$: 
${}^3\eps_{bcd}\equiv n^a\eps_{abcd}$.
Now using the axisymmetry of the $\chi^{\pm}$,
and writing $U^a$ as $U^a = v^a/|v|$ with angular velocity $\Omega$
as given above in (\ref{tmp1}), we have (using $\de\xi^a=\de\phi^a=0$) 
\bea
\overline{dX}&=&-{}^3\eps 2 n n_e \Omega
\de(\mu U_r) \varphi^{[e}\xi^{r]}/|v|\nonumber\\
&=& {}^3\eps (p+\rho)\Omega (n_e\xi^e)\de(U_r) \varphi^{r}/|v|\nonumber\\
&=&-{}^3\eps\Omega (p+\rho)\de(U_r)\varphi^r\nonumber\\
&=&\Omega \de ({}^3\eps n_a{T^a}_b \varphi^b)\nonumber\\
&=&-\Omega \de \overline{J}
\eea
where $\overline{J}$ is the pullback of $J_{abc}$ to $\Sigma$. 
Therefore (\ref{tmp1}) now matches (\ref{tmp2}) and so the first law 
(\ref{fl}) now agrees with the first law given in (\ref{bchfl}).

\section*{Acknowledgements}
This research was supported in part by NSF grant PHY-95-14726 to the 
University of Chicago. This work is presented as a thesis to the 
Department of Physics, The University of Chicago, in partial 
fulfillment of the requirements for the Ph.D. degree. I am very grateful 
for Bob Wald's expert guidance throughout this project. I also wish to thank 
Bob Geroch, Eanna Flanagan and Shyan-Ming Perng for helpful discussions.

\section*{Appendix: The Chandrasekhar-Ferrari conserved current}
The symplectic form $\om(\phi,\de_1\phi,\de_2\phi)$ defined in (\ref{omega})
is closed when $\de_{1,2}\phi$ satisfy the
linearised equations. Its dual $\omega^d(\phi,\de_1\phi,\de_2\phi)$, defined 
by $\om_{abc}=\omega^d\eps_{dabc}$,
is therefore a covariantly conserved current for the 
Einstein-perfect fluid system. 
Chandrasekhar and Ferrari \cite{CF} have, from first principles,
also derived a conserved current, ${\cal E}^{a}(\phi,\de\phi)$, for 
the Einstein-perfect
fluid system. Their current is quadratic in the (complex) perturbations 
$\de\phi$, and is restricted to the case where 
$\phi$ is a static axisymmetric solution, and $\de\phi$ is a ``polar"
(even parity) perturbation with harmonic time dependence (we will define
this below). 
We now show the equivalence of the $\omega^a(\phi,\de\phi,\de\phi^*)$ 
and ${\cal E}^{a}$ for the Einstein-perfect fluid system.
This calculation is the analogue
for the Einstein-perfect fluid system of the calculation by Burnett and
Wald \cite{BW} for the Einstein-Maxwell system. 

We start by choosing the Lagrangian for the Einstein-perfect fluid system to be 
\be
\bL_{pqrs} = \eps_{pqrs}[ - \frac{1}{4} R + P(m,\sigma) ],
\ee
where we have set the constant in front of the Ricci scalar to give the
field equations in \cite{CF}, and used Schutz's velocity-potential
representation, with Lagrangian (\ref{Lepf}).
The symplectic potential, $\syp$, arising from this Lagrangian is 
(after substituting (\ref{Udef}) where applicable),
\bea
\syp_{pqr}=-\frac{1}{4}\eps_{apqr}(\del^b{\gamma^a}_b-\del^a\gamma)
-nU^a\eps_{apqr}(\de\Phi+\alpha\de\beta+\theta\de s),
\label{thetaepf}
\eea
where $\gamma_{ab} \equiv \de g_{ab}$ and $\gamma\equiv g^{ab}\gamma_{ab}$.
The resulting presymplectic form is (from (\ref{omega}))
\bea
\om_{pqr}(\phi,\de_1\phi,\de_2\phi)
&=&-\frac{1}{8}\eps_{apqr}[({\gamma_2}^{cd}-g^{cd}{\gamma_2})
\del^a{\gamma_1}_{cd} - (2{\gamma_2}^{cd}
-{\gamma_2} g^{cd})\del_c{{\gamma_1}_{d}}^a
+{\gamma_2}^{ad}\del_d{\gamma_1}]\nonumber\\
&& - \de_2(\eps_{apqr}nU^a)(\de_1\Phi+\alpha\de_1\beta+\theta\de_1 s)
\nonumber\\
&&- \eps_{apqr}nU^a(\de_2\alpha\de_1\beta + \de_2\theta\de_1 s) 
-(1\leftrightarrow 2).
\label{epfsf}
\eea
This form is dual to a generally conserved current: it can be
shown \cite{LW} that for $\omega^a$ defined above, we have (for perturbations
$\de_1\phi$ and $\de_2\phi$ satisfying the linearised field equations), 
\be
\del_a \omega^a = 0.
\ee

We now relate this conserved current to the current presented in \cite{CF}, 
by fixing a coordinate system with derivative operator $\dee_a$, 
and writing the volume element $\eps$
in terms of the coordinate volume element $\bf e$ of this system --
\be
\eps_{abcd} = \sqrt{-g} {\bf e}_{abcd}
\ee
-- then the vector field $w^a$ defined by $\om_{pqr} = w^a{\bf e}_{apqr}$ 
is conserved in the sense $\dee_a w^a = 0$. If we 
follow Chandrasekhar and Ferrari \cite{CF} and specialise to the case where the 
background spacetime is static (with static killing field $t^a$)
and axisymmetric (with axial killing field $\varphi^a$), and the perturbations
are time and angle-dependent only ``harmonically'' (that is, there are
constants $\sigma$ and $\omega$ such that 
\bea
\Liet\de\eta &=&i\sigma\de\eta \nonumber\\
{\cal L}_{\varphi} \de\eta&=&i\omega\de\eta,
\label{harm}
\eea
for all the dynamical fields $\eta$) then (following \cite{BW}) it's easy to
see that for complex $\de\phi$, $w^t(\phi,\de\phi,\de\phi^*)$ and 
$w^{\varphi}(\phi,\de\phi,\de\phi^*)$ are independently 
conserved: $\dee_tw^t+ \dee_{\varphi}w^{\varphi} = 0$. We can therefore
restrict our attention to the vector components $(w^2, w^3)$.
Moreover, (\ref{harm}) allows us to substitute the variations of the
fluid potentials $\de (\Phi,\beta,s)$ 
for variations of their time derivatives: 
we do this and (recalling (\ref{Udef}))
find
\be
w^a = w^a_{gr} - \de_2(\sqrt{-g}nU^a)\frac{1}{i\sigma}\de_1(\mu t^b U_b)
-\sqrt{-g}\frac{1}{i\sigma}nU^a[\de_2(t\cdot\del\alpha)
\de_1(t\cdot\del\beta)+\de_2(t\cdot\del\theta)\de_1(t\cdot\del s)] 
-(1\leftrightarrow 2),
\label{epfsfharm}
\ee
where we labelled the contribution from the first 
two lines of (\ref{epfsf}) by $w^a_{gr}$.

Our aim is now to show the equality of 
$(w^2(\phi,\de\phi,\de\phi^*), w^3(\phi,\de\phi,\de\phi^*))$ and
$({\cal E}^2, {\cal E}^3)$. To do this we first specialise the background
and perturbations in $w^a$ to those used by Chandrasekhar and Ferrari. 
In the coordinates given in \cite{CF} the metric is written
\be
g_{ab} = -e^{2\nu}dt_adt_b+e^{2\psi}d\varphi_ad\varphi_b+e^{2\mu_2}dx^2_adx^2_b
+e^{2\mu_3}dx^3_adx^3_b,
\label{cfg}
\ee
and the nonvanishing (polar) metric perturbations are taken to be
\bea
{\gamma_1}_{tt}& =& -2 e^{2\nu}\de\nu\nonumber\\
{\gamma_1}_{\varphi\varphi}&=& 2 e^{2\psi}\de\psi\nonumber\\
{\gamma_1}_{22}& =& 2 e^{2\mu_2}\de\mu_2\nonumber\\
{\gamma_1}_{33}& =& 2 e^{2\mu_3}\de\mu_3,
\eea
We then set ${\gamma_2}_{ab} = {\gamma_1}_{ab}^*$, where the perturbed 
functions, $\de\nu$, etc. are complex, but the unperturbed functions are real.

A direct substitution of these perturbations into $w^a_{gr}$ yields
the result already known from \cite{BW};
\bea
w_{gr}^2 &=&-\frac{1}{2}e^{\nu+\psi-\mu_2+\mu_3}
[\de\nu_{,2}\de(\psi+\mu_3)^* \nonumber\\
&&\de\psi_{,2}\de(\nu+\mu_3)^*+\de\mu_{3,2}\de(\nu+\psi)^*\nonumber\\
&&+\nu_{,2}\de(\psi+\mu_3)^*\de(\nu-\mu_2) \nonumber\\
&&+\psi_{,2}\de(\nu+\mu_3)^*\de(\psi-\mu_2) \nonumber\\
&&+\mu_{3,2}\de(\mu+\psi)^*\de(\mu_3-\mu_2)] - c.c.
\label{wgr}
\eea

We now turn to the fluid contributions $w^a_f$ to the conserved current,
defined by $w^a_f = w^a-w^a_g$. We set the four-velocity of the
background to be $U^a = e^{-\nu}t^a$ and (following \cite{CF})
and denote the perturbations of the {\em orthonormal frame} 
components of $U^a$ by $i\sigma\xi^a$: $\de U_{\hat{a}} = i\sigma \xi_a$. 
We then find the `2'-component of $w_f^a$ given by
\bea
w_f^2&=& -dx^2_a\frac{1}{i\sigma}[\de_2(e^{\nu+\psi+\mu_2+\mu_3}nU^a)
\de_1(\mu t^cU_c)] - (1\leftrightarrow 2) \nonumber\\
&=&
\frac{1}{i\sigma}e^{2\nu+\psi+\mu_3}n\de_2(U^{\hat{2}})(\de_1\mu 
+\mu\de_1\nu - \mu \de_1U_{\hat{0}})-(1\leftrightarrow 2) \nonumber\\
&=&
-e^{2\nu+\psi+\mu_3}(n\de\mu + n\mu\de\nu) \xi^*_2 - c.c.,
\label{ftmp}
\eea
where we have set $\de_1=\de$ and $\de_2=\de^*$, and used the 
result (see \cite{CF}) that $\de U_{\hat{0}} = 0$.
We can also 
put $n\de\mu=\de p + nT\de s$, and bearing in mind that $\de s$ must
also have harmonic time dependence, we can write 
\bea
i\sigma n T \de s &=& nT t\cdot\del\de s\nonumber\\
&=& nT e^{\nu}U\cdot\del \de s\nonumber\\
&=& nT e^{\nu}(\de(U\cdot\del s)-\de(U)\cdot\del s).
\label{deS}
\eea
Referring to (\ref{vortfe}) we see that
the first term on the right side of (\ref{deS}) vanishes whenever
the perturbation satisfies the linearised equations. Since the
background is vortex-free, we see that the second term 
also vanishes as a consequence of (\ref{vortfe}).
Adding the resulting fluid contribution to the gravitational terms
(\ref{wgr}) yields 
\bea
w^2 &=&-\frac{1}{2}e^{\nu+\psi-\mu_2+\mu_3}
[\de\nu_{,2}\de(\psi+\mu_3)^* \nonumber\\
&&\de\psi_{,2}\de(\nu+\mu_3)^*+\de\mu_{3,2}\de(\nu+\psi)^*\nonumber\\
&&+\nu_{,2}\de(\psi+\mu_3)^*\de(\nu-\mu_2) \nonumber\\
&&+\psi_{,2}\de(\nu+\mu_3)^*\de(\psi-\mu_2) \nonumber\\
&&+\mu_{3,2}\de(\nu^*+\de\psi^*)\de(\mu_3-\mu_2)]\nonumber\\
&&- e^{2\nu+\psi+\mu_3}(\de p + (p+\rho)\de\nu)  \xi^*_2 - c.c.
\label{w2}
\eea

Now using the appropriate linear combinations of the 
linearised Einstein constraint -- 
\be
\de(\psi_+\mu_3)_{,2} +\psi_{,2}\de(\psi-\mu_2)
+\mu_{3,2}\de(\mu_3-\mu_2)-\nu_{,2}\de(\psi+\mu_3)=2 e^{\nu+\mu_2}(\rho+p)
\xi^2 
\ee
-- to replace the third, fourth and fifth lines of (\ref{w2}) we get 
\bea
w^2&=&-\frac{1}{2}e^{\nu+\psi+\mu_3-\mu_2}\{\de\nu_{,2}\de(\psi+\mu_3)^*
+\de\mu^*(\de\psi+\de\mu_3)_{,2} \nonumber\\
&&-[\de\psi,\de \psi^*]_{,2}-[\de\mu_3,\de \mu_3^*]_{,2} \nonumber\\
&&+2e^{\nu+\mu_2}((\rho+p)\de(\psi+\mu_3-\mu_2)^*-\de p^*)\xi_2\}- c.c.
\eea
where, we define $[A,A^*]_{,i} \equiv A_{,i}A^*-A A^*_{,i}$. 
This is seen to agree (up to an overall constant) with ${\cal E}^2$
of the conserved current in \cite{CF}. A similar calculation for $w^3$
yields ${\cal E}^3$ (which is obtained from ${\cal E}^2$ by interchanging
$2\leftrightarrow 3$), and so we find $(w^2,w^3)=({\cal E}^2,{\cal E}^3)$, 
and our symplectic current $w^a$ for the Einstein-perfect fluid system
agrees with the Chandrasekhar-Ferrari current for this system. 

We make two final comments. Firstly, from the comment following
Eq.(\ref{Udef}), we know that every configuration of the 
physical fields of a perfect fluid has a corresponding equivalence class 
of configurations of the dynamical fields, and as a consequence, 
every perturbation of the physical fields has a corresponding perturbation
of the dynamical fields. Now, two distinct perturbations of the physical
fields off the same background (physical field) configuration will each select 
a corresponding perturbation of the dynamical fields. The background 
{\em dynamical} field configuration for each of these perturbations 
will certainly lie within the equivalence class corresponding to the given 
background physical field 
configuration: however, in general, these background dynamical field 
configurations will 
be {\em distinct} elements of this equivalence class. In using symplectic
methods to derive ${\cal E}^a$ we have implicitly restricted ourselves to those
pairs of perturbations of the physical fields where the corresponding pairs
of dynamical field perturbations $(\de_1\phi,\de_2\phi)$ 
have {\em identical} background configurations. In fact, as we have seen above, 
the resulting conserved current agrees with the Chandrasekhar - Ferrari
current for {\em all} pairs of perturbations of the physical fields, 
not just those restricted in this way.

Secondly, we notice from (\ref{epfsfharm}) that as long as the $U^a$ of 
the background solution
lies in a plane tangent to the subspace spanned by $t^a$ and $\varphi^a$, 
the last term in (\ref{epfsfharm}) vanishes for the components 
of interest. This in turn yields a conserved current $(w^2,w^3)$ which only
depends on perturbations of the {\em physical} fields, without the 
explicit appearance
of the fluid potentials, for any stationary background configuration
in which the fluid velocity is tangent to the $t-\varphi$ subspaces. 
Of course, we know that $\omega^a$ is a conserved current off {\em any} 
background; this observation suggests only that a current similar 
in style to that presented by Chandrasekhar and Ferrari also exists 
for a background with a fluid in circular motion, as well as the static 
case considered in \cite{CF}.

\end{document}